\documentclass{IEEEojcsys}

\usepackage[colorlinks,urlcolor=blue,linkcolor=blue,citecolor=blue]{hyperref}

\usepackage{array}

\usepackage{graphicx}
\usepackage{amsmath}
\usepackage{amsfonts}
\usepackage{bm}
\usepackage{soul,color}
\usepackage{subcaption}
\usepackage{multirow}

\usepackage{algorithm}
\usepackage{algorithmic}

\jvol{00}
\jnum{XX}
\paper{1234567}
\pubyear{2021}
\receiveddate{XX September 2021}
\accepteddate{XX October 2021}
\publisheddate{XX November 2021}
\currentdate{XX November 2021}
\doiinfo{OJCSYS.2021.Doi Number}

\begin{document}

\sptitle{Article Category}

\title{VisioPath: Vision-Language Enhanced Model Predictive Control for Safe Autonomous Navigation in Mixed Traffic} 

\editor{This paper was recommended by Associate Editor F. A. Author.}

\author{S. WANG\affilmark{1}}
\author{P. TYPALDOS\affilmark{2}}
\author{C. LI\affilmark{3}}
\author{A. A. MALIKOPOULOS\affilmark{1,2}}

\affil{System Engineering Program, Cornell University, Ithaca, NY 14850, USA} 
\affil{School of Civil and Environmental Engineering, Cornell University, Ithaca, NY 14850, USA} 
\affil{School of Electrical and Computer Engineering, Cornell University, Ithaca, NY 14850, USA} 

\corresp{CORRESPONDING AUTHOR: S. Wang (e-mail: \href{mailto:sw997@cornell.edu}{sw997@cornell.edu})}
\authornote{This research was supported in part by NSF under Grants CNS-2401007, CMMI-2348381, IIS-2415478, and in part by MathWorks.}

\markboth{PREPARATION OF PAPERS FOR IEEE OPEN JOURNAL OF CONTROL SYSTEMS}{S. WANG {\itshape ET AL}.}

\begin{abstract}
In this paper, we introduce \textit{VisioPath}, a novel framework combining vision-language models (VLMs) with model predictive control (MPC) to enable safe autonomous driving in dynamic traffic environments. The proposed approach leverages a bird’s-eye view video processing pipeline and zero-shot VLM capabilities to obtain structured information about surrounding vehicles, including their positions, dimensions, and velocities. Using this rich perception output, we construct elliptical collision-avoidance potential fields around other traffic participants, which are seamlessly integrated into a finite-horizon optimal control problem for trajectory planning. The resulting trajectory optimization is solved via differential dynamic programming with an adaptive regularization scheme and is embedded in an event-triggered MPC loop. To ensure collision-free motion, a safety verification layer is incorporated in the framework that provides an assessment of potential unsafe trajectories. Extensive simulations in Simulation of Urban Mobility (SUMO) demonstrate that \textit{VisioPath} outperforms conventional MPC baselines across multiple metrics. By combining modern AI-driven perception with the rigorous foundation of optimal control, \textit{VisioPath} represents a significant step forward in safe trajectory planning for complex traffic systems.
\end{abstract}

\begin{IEEEkeywords}
Vision-language models (VLMs), AI-driven perception, mixed traffic, safe autonomous navigation.
\end{IEEEkeywords}

\maketitle

\section{INTRODUCTION}
Recent advancements in artificial intelligence (AI) and sensor technologies have accelerated the development of autonomous vehicles (AVs). Despite significant progress, deploying AVs remains challenging due to the complexity and unpredictability of mixed-traffic environments. Existing trajectory planning methods for AVs include Hamiltonian control theory \cite{malikopoulos2021optimal, malikopoulos2018decentralized, mahbub2020sae-1}, reinforcement learning (RL) \cite{kiran2021deep, aradi2020survey, nakka2022multi}, and model predictive control (MPC) \cite{Le2023ACC, le2024controller, guanetti2018control, typaldos2022optimization}. 

More recently, vision-language models (VLMs) and large language models (LLMs) have emerged as potential solutions for directly extracting contextual information and generating trajectory recommendations. However, these models lack safety guarantees and may suffer from inconsistent outputs due to hallucination problems, making them unreliable as solutions for safety-critical applications \cite{li2025fine,zhou2024vision, yang2023llm4drive}.

To address these limitations, we introduce \textit{VisioPath}, a framework that integrates VLMs with differential dynamic programming (DDP) \cite{murray1979constrained} within an MPC structure. \textit{VisioPath} leverages bird's-eye view (BEV) perception combined with zero-shot VLM capabilities to rapidly extract structured vehicle information. This data facilitates the real-time construction of elliptical collision-avoidance fields. The optimization employs constrained DDP with adaptive regularization, embedded in an event-triggered MPC loop for efficient, dynamic replanning.

To validate the performance of our framework, we conducted extensive simulations in SUMO which demonstrate \textit{VisioPath}'s superior performance in travel efficiency, trajectory optimality, and safety margins compared to traditional MPC methods. Our approach combines the visual understanding and reasoning of VLMs with the formal guarantees of optimization-based planning, creating a robust solution for autonomous navigation in complex environments.

\section{CONTRIBUTION}
We develop \textit{VisioPath}, a hybrid control framework integrating zero-shot VLMs with DDP. Unlike existing approaches, our system combines advanced AI perception with the mathematical rigor of optimal control theory.
Moreover, we implement a model-free BEV preprocessing pipeline that extracts vehicle bounding boxes and speeds in approximately 20 ms per frame without requiring heavy neural network computation, enabling real-time operation with sufficient resources for downstream planning. Unlike the related approach in \cite{tian2024drivevlm}, we design a safety verification system utilizing elliptical collision-avoidance fields that evaluates trajectories against multiple metrics including collision proximity, time-to-collision thresholds, and road boundary compliance. Finally, we conduct extensive simulations in SUMO to demonstrate that \textit{VisioPath} significantly outperforms conventional MPC approaches in travel efficiency, computational load, and safety margins across diverse high-traffic scenarios.

\section{RELATED WORK}

\subsection{VLM/Perception in Autonomous Driving}

Recent advancements in VLMs are significantly influencing autonomous driving systems by integrating visual and linguistic information to enhance capabilities ranging from scene understanding to decision-making and planning. Tian et al. introduced \textit{DriveVLM}, which leverages VLMs for improved scene understanding and hierarchical planning, effectively handling complex urban scenarios \cite{tian2024drivevlm}. Similarly, Guo et al. proposed \textit{Co-driver}, a VLM-based assistant designed to emulate human-like driving behaviors by adjusting strategies based on real-time scene understanding, thereby enhancing adaptability in complex road conditions, such as under different weather or light conditions \cite{guo2024co}.

Alongside system development, significant efforts also focus on evaluating and surveying the role of these models. To address crucial safety and reliability concerns, Xing et al. developed \textit{AutoTrust}, a comprehensive benchmark for evaluating VLM trustworthiness aspects such as safety, robustness, privacy, and fairness \cite{xing2024autotrust}. Zhou et al. conducted an extensive survey covering applications in perception, planning, decision-making, and data generation, identifying current research trends and future directions \cite{zhou2024vision}. Complementarily, Fourati, et al. reviewed the potential and deployment of Cross-modal Language Models, discussing architectures and frameworks that emphasize integrating multimodal sensory inputs with language understanding to enhance decision-making and control in autonomous vehicles \cite{fourati2024xlm}. Choudhary et al. \cite{choudhary2024talk2bev} introduced a BEV interface specifically designed for autonomous driving. These developments show a growing consensus that VLMs when properly integrated, can provide high-level understanding, reasoning, and decision support that complement the strengths of traditional perception and control modules. Inspired by these efforts, in this paper, we integrate BEV precognition with VLMs' reasoning ability to generate a recommended initial trajectory.

In our BEV pre‑processing, we generate at most 15 proposals per frame in approximately $20$\, ms by combining a fixed homography warp, illumination normalization, and motion‑guided connected‑component analysis reasoning that is the focus of this work—avoiding any time-consuming neural-network inference and leaving the computational budget for the downstream vision–language. If higher detection fidelity is desired, the same interface can ingest proposals from common detectors such as Faster R‑CNN~\cite{ren2015faster}, the real‑time YOLOv4 family~\cite{bochkovskiy2020yolov4}, or monocular BEV networks like Lift‑Splat‑Shoot~\cite{philion2020lift}, without requiring changes to the MPC or safety layers.

\subsection{MPC/DDP in Autonomous Driving}
DDP, introduced in \cite{mayne1970} and later extended by \cite{murray1979constrained, murray1984differential}, is an iterative algorithm that progressively improves trajectories until converging to an optimal control solution. Each iteration computes a quadratic-linear approximation of the recursive Bellman equation around the current trajectory.  
DDP has been applied in several areas, including robotics, unmanned aerial vehicles (UAV), connected and automated vehicles, multi-agent systems, and recently, urban air mobility. Specifically, the authors in \cite{tassa2014control} address the challenge of enforcing actuator and torque limits while retaining the computational efficiency of DDP. Their main application is real-time whole-body humanoid control: they demonstrate the algorithm on the humanoid robot, showcasing dynamically feasible motions like balancing and reaching while respecting strict control bounds. The authors in \cite{xie2017differential} extended DDP to handle arbitrary nonlinear constraints on both states and controls. The approach was demonstrated using UAVs that maneuver around obstacles, verifying that the DDP can maintain feasibility even with dynamic constraints. 

A control framework of decentralized multi-decision-makers that combine DDP with the alternating direction method of multipliers algorithm to address large-scale problems involving hundreds of agents was introduced in \cite{saravanos2023distributed}. The framework was demonstrated in large-scale simulations by coordinating 1024 cars and validated the approach with hardware experiments on a real multi-robot platform. In a different context, the authors in \cite{typaldos2023modified} focused on the green light optimal speed advisory problem, considering adaptive signals, and achieved real-time numerical solutions for vehicles approaching traffic signals using discrete DDP and DDP. 

In our previous work \cite{wang2025corra}, we introduced a hybrid framework that leverages LLMs' reasoning capabilities to dynamically generate sigmoid-based safety boundaries around obstacle vehicles. The approach integrates with DDP within an MPC framework, treating safety boundaries as hard constraints rather than soft penalties, enabling efficient real-time trajectory planning in complex mixed-autonomy traffic environments.

In this paper,  we demonstrate that with VLM perception understanding, an event-triggered DDP can be effectively deployed in a loop for highly dynamic environments. It is worth noting that extending DDP to handle complex constraints and uncertainties is an active research topic. Traditional DDP does not natively handle state/input constraints or chance constraints, but recent efforts have made progress. One approach is to formulate constraints as soft penalties or incorporate them via augmented Lagrangian methods \cite{howell2022trajectory, jallet2022constrained}. However, naive penalty approaches can suffer from convergence issues or suboptimal solutions in practice. Alcan et al. developed Safe-CDDP to address safety-critical constraints, which introduces chance constraints into the DDP framework to ensure a desired probability of safety under modeling uncertainties \cite{alcan2022differential}. Our \textit{VisioPath} framework similarly emphasizes constraint satisfaction and safety: we embed collision-avoidance constraints via potential fields and perform explicit trajectory validation post-optimization, drawing inspiration from these constrained DDP strategies to guarantee safety without compromising performance.

\section{PROBLEM FORMULATION}
\label{sec:problem_formulation}

In this section, we present the mathematical formulation of our proposed hybrid framework that combines MPC with VLMs for safe trajectory planning in mixed autonomy traffic environments.

\begin{figure*}[ht]
    \centering
    \includegraphics[width=\linewidth]{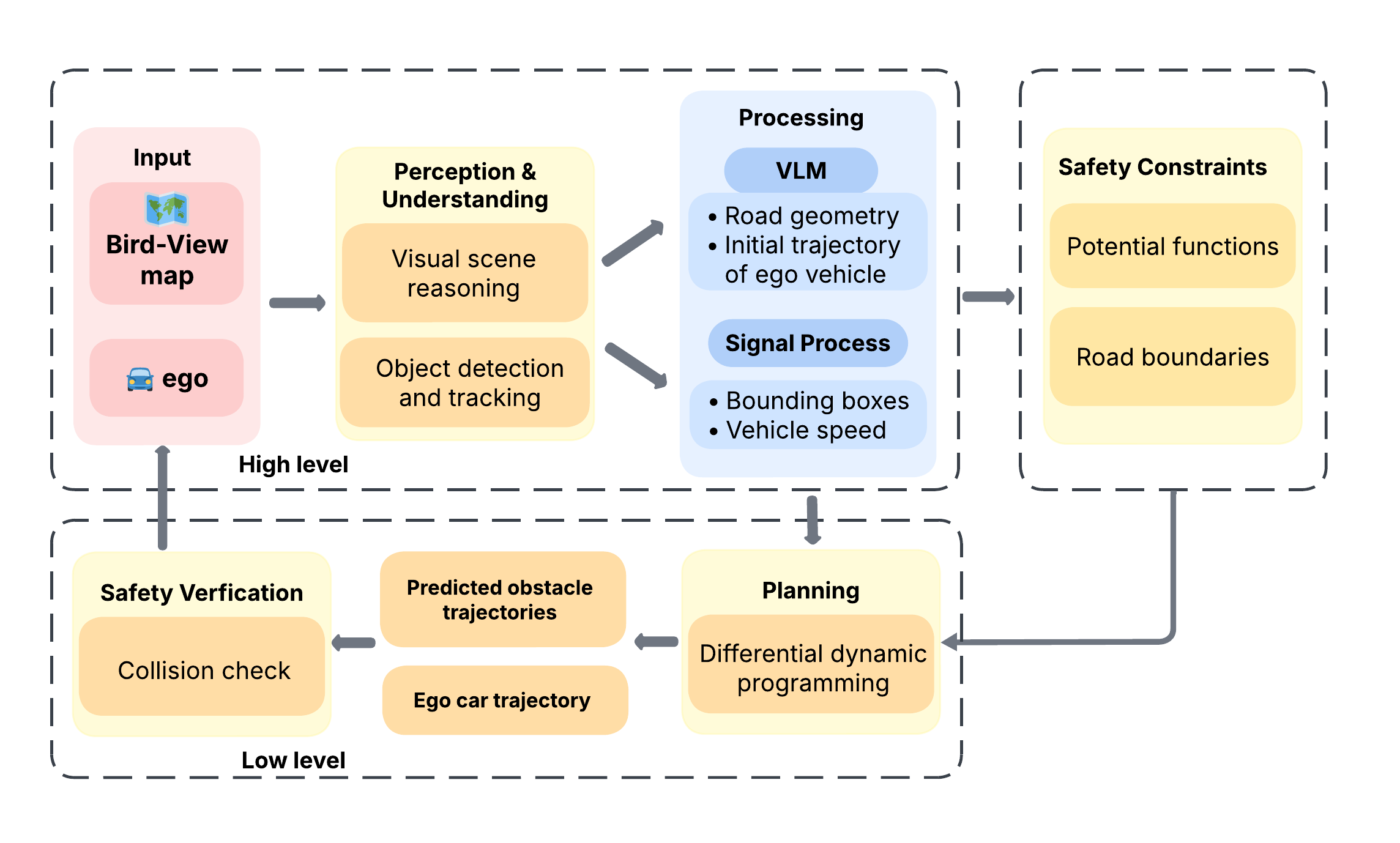}
    \caption{Overall pipeline framework of our method, \textit{VisioPath}.}
    \label{fig:overall_pipeline}
\end{figure*}

\subsection{System Dynamics and Constraints}
We consider an AV that navigates a multilane road segment with human-driven vehicles and other AVs. The states of the controlled AV (Ego) at discrete time step $k$ is denoted by $\bm{x}_k \in \mathbb{R}^n$, which includes longitudinal and lateral positions and velocities. The control inputs are represented by $\bm{u}_k \in \mathbb{R}^m$, which includes longitudinal and lateral accelerations.

The discrete-time dynamics of the AV can be expressed as:
\begin{equation} \label{eq:states_gen}
\bm{x}_{k+1} = \bm{f}(\bm{x}_k, \bm{u}_k),
\end{equation}
where $\bm{x}_k = [x_k, y_k, v_{x,k}, v_{y,k}]^T$ is the states vector and $\bm{u}_k = [u_{x,k}, u_{y,k}]^T$ is the controls vector. The state equations (1) can be described in matrix form as follows:

\begin{equation} \label{eq:states}
    \begin{bmatrix}
        x_{k+1} \\
        y_{k+1} \\
        v_{x,k+1} \\
        v_{y,k+1}
    \end{bmatrix} = 
    \begin{bmatrix}
        1 & 0 & T & 0 \\ 
        0 & 1 & 0 & T \\
        0 & 0 & 1 & 0 \\
        0 & 0 & 0 & 1
    \end{bmatrix}
    \begin{bmatrix}
        x_{k} \\
        y_{k} \\
        v_{x,k} \\
        v_{y,k}
    \end{bmatrix} + \begin{bmatrix}
        \tfrac{1}{2} T^2 & 0 \\
        0 & \tfrac{1}{2} T^2 \\
        T & 0 \\
        0 & T
    \end{bmatrix}
    \begin{bmatrix}
        u_{x,k} \\
        u_{y,k} \\
    \end{bmatrix},
\end{equation}
where the state variables $x_k, y_k, v_{x,k}, v_{y,k}$ represent the longitudinal and lateral positions and speeds at discrete time step $k$, respectively, while $u_{x,k}, u_{y,k}$ are the control variables reflecting on the longitudinal and lateral accelerations. Note that the control variables are kept constant for the duration $T$ of each time step $k$.

The control variables $\bm{u}_k = [u_{x,k}, u_{y,k}]^T$ are bounded according to the specifications and restrictions of the vehicle as follows:
\begin{equation}
    \bm{u}^{L}(\bm{x}_k) \leq \bm{u}_{k} \leq \bm{u}^{U}(\bm{x}_k),
\end{equation}
where $\bm{u}^{L}(\bm{x}_k)=[u_{x,L}(\bm{x}_k), u_{y,L}(\bm{x}_k)]$ and $\bm{u}^{U}(\bm{x}_k)=[u_{x,U}(\bm{x}_k), u_{y,U}(\bm{x}_k)]$; in the longitudinal direction, the upper bound in (3) is constant, i.e., $u_{x,U}(\bm{x}_k) = u_x^{\max}$, and reflects the acceleration capabilities of the vehicle. On the other hand, the lower bound is designed appropriately as a state-dependent bound
\begin{equation} \label{eq: bound_y_low}
    u_{x,L}(\bm{x}_k) = \max\left\{-\frac{1}{T}v_{x,k}, u_x^{\min}\right\}.
\end{equation}

The last equation guarantees that the vehicle does not reach negative longitudinal speed values and that its lower value is greater than or equal to a constant minimum value $u_x^{\min}$, which represents the maximum deceleration that the vehicle can achieve.

The lateral acceleration bounds ensure that the vehicle will not violate the road boundaries, which, in this work, are straight lines. Thus, the derived upper and lower bounds in (3) are given by

\begin{equation}
    u_{y,U}(\bm{x}_k) = \dfrac{2(\tilde{y}_l - y_k - v_{y,k}\cdot T)}{T^2},
\end{equation}
\begin{equation}
    u_{y,L}(\bm{x}_k) = \dfrac{2(\tilde{y}_r - y_k - v_{y,k}\cdot T)}{T^2},
\end{equation}
where \(\tilde{y}_l\) and \(\tilde{y}_r\) are the lateral positions of the left and right boundaries, i.e., \(\tilde{y}_l = r_w - \frac{w_{\text{Ego}}}{2}\) and \(\tilde{y}_r = \frac{w_{\text{Ego}}}{2}\), respectively, with \(r_w\) being the width of the road and \(w_{\text{Ego}}\) the width of the Ego vehicle.

\begin{figure*}[ht]
    \centering
    \includegraphics[width=\linewidth]{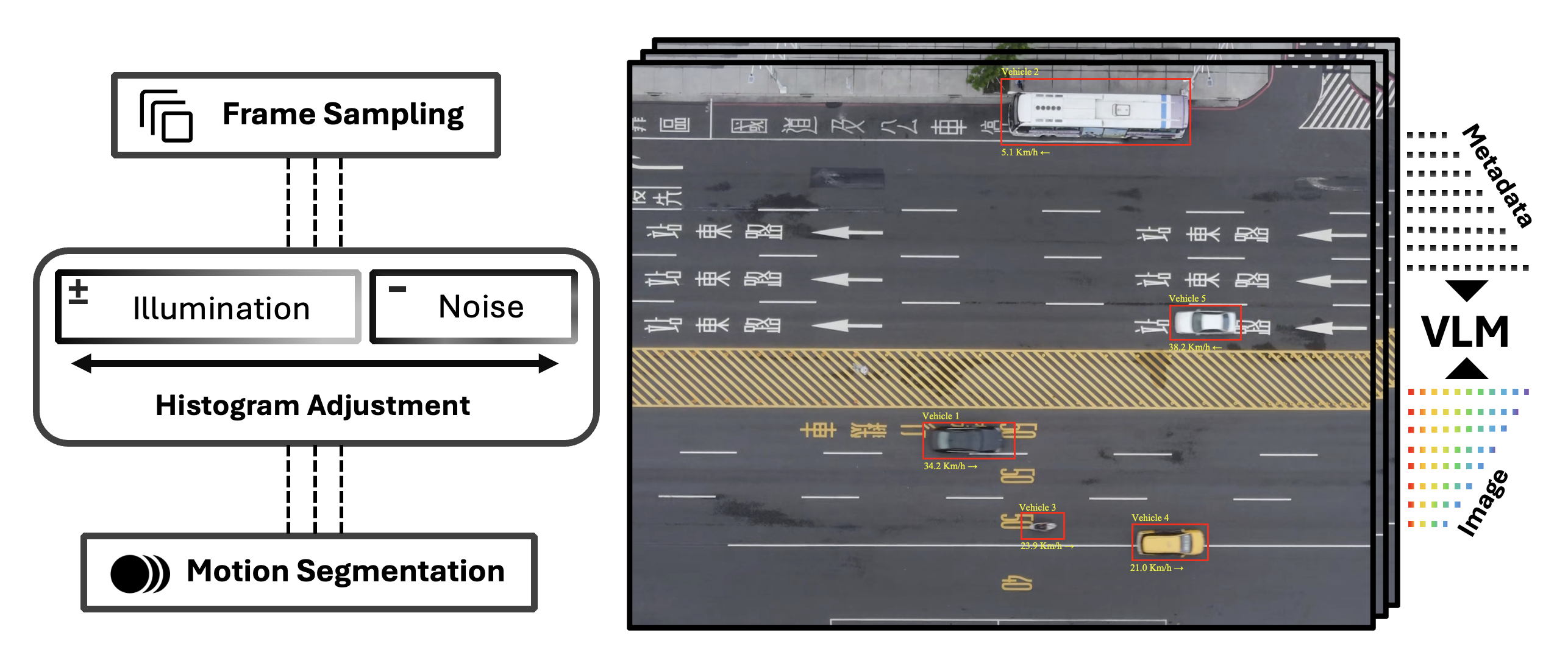}
    \caption{Pipeline of bird's-eye-view video pre-processing.}
    \label{fig:birdview_preprocess}
\end{figure*}

\subsection{BEV Video Pre‑processing}\label{subsec:bev_preproc}

The goal of the BEV pre‑processing module is to transform raw video frames from a traffic camera into a small, structured list of vehicle bounding boxes with metric coordinates and velocity estimates.  
This information is computationally feasible for real-time use and is rich enough as input for the downstream vision–language module and optimization layer. 

The challenges we face include variations, camera distortions, background clutter, and noise that would confuse any automated analysis \cite{jahne2005digital}. We followed a five-stage pipeline to address these issues. The proposed pipeline, illustrated in Fig.~\ref{fig:birdview_preprocess}, is fully deterministic and consists of the following five stages:

\begin{enumerate}

\item \textbf{Frame acquisition and temporal sub‑sampling.}
To reduce the computational load while maintaining sufficient temporal resolution for vehicle tracking, we select the frame rate to $f_{s}=10$\, Hz (10 frames per second). This is sufficient for urban traffic speeds without losing important motion information since the urban traffic moves relatively slowly compared to the camera's capture rate ($f_{s}=30$\, Hz). Therefore, raw RGB frames $\mathbf{I}_{k}^{\text{raw}}\!\in\!\mathbb{R}^{H\times W\times 3}$ are sampled at $f_{s}=10$\,Hz, where each frame contains height H, width W, and 3 color channels. This sub-sampling reduces our computational burden.

\item \textbf{Metric rectification by a homography matrix.}  
To convert the camera's angled perspective view into an accurate true bird's eye view with real-world measurements, we use a mathematical transformation called a \emph{homography} matrix $\mathbf{H}$ (a $3{\times}3$ projective transformation) that maps pixel coordinates $(u,v)$ in the original camera image to real-world ground coordinates $(\tilde{x},\tilde{y})$ measured in meters.
We obtain $\mathbf{H}$ by performing a one-time calibration by placing a chessboard calibration pattern on the road and solving a standard least‑squares problem that matches corner pixels to known world points.
      At run time we warp every subsampled frame:
      \[
        \tilde{\mathbf{p}} = \mathbf{H}\,[u\,v\,1]^{\!\top},
        \qquad
        \tilde{\mathbf I}_k = \mathcal W\!\bigl(\mathbf{I}^{\text{raw}}_k,\mathbf H\bigr),
      \]
      where $\mathcal W(\cdot)$ carries out inverse mapping plus bilinear interpolation.  
      The result $\tilde{\mathbf I}_k$ is a true top‑down view in \emph{meters per pixel}; the axes are aligned with the road, and the following geometric checks are greatly simplified.

\item \textbf{Illumination normalization and noise suppression.}  
      To eliminate lighting variations and sensor noise that would interfere with accurate vehicle detection, we therefore apply (i) a Retinex‑based adaptive histogram equalization $\mathcal R(\cdot)$ to stabilize brightness and (ii) a bilateral filter $\mathcal B(\cdot)$ to smooth pixel noise while keeping edge locations sharp \cite{liu2021retinex, tomasi1998bilateral}:
      \[
        \mathbf{I}_k = \mathcal B\!\bigl(\mathcal R(\tilde{\mathbf I}_k)\bigr).
      \]

\item \textbf{Motion‑guided foreground segmentation.}  
      Let $\bar{\mathbf I}_k$ be an exponentially weighted moving average of the past frames—the \emph{background model}. To identify which parts of the image contain moving vehicles versus static background elements, we maintain a background model $\bar{\mathbf I}_k$ and then identify foreground pixels using two criteria.
      Pixels whose intensity deviates by more than a threshold $\tau_{\text{bg}}$ from that background, \emph{or} whose optical‑flow magnitude exceeds $\tau_{\text{flow}}$, are flagged as belonging to moving objects:
      \[
        \mathcal M_k=\bigl\{\mathbf x\;|\;
          |\mathbf I_k(\mathbf x)-\bar{\mathbf I}_k(\mathbf x)|>\tau_{\text{bg}}
          \,\vee\,
          \|\mathbf F_k(\mathbf x)\|>\tau_{\text{flow}}
        \bigr\}.
      \]
      Connected‑component analysis together with a morphological closing operation turns this binary mask into coarse “blobs,” each expected to contain exactly one vehicle.

\item \textbf{Bounding‑box generation and selection.}  
      For every blob we compute the smallest axis‑aligned rectangle that fully encloses it, then expand the sides by $20\%$ as a safety margin.  
      Each rectangle is cropped out of $\mathbf I_k$, resized to $224{\times}224$\,px, and stored as a \emph{candidate bounding box}
      \[
        c_{i,k}=\bigl(\mathbf I^{\text{crop}}_{i,k},\;
                      \mathbf p^{\text{img}}_{i,k},\;
                      s_{i,k}\bigr),
      \]
      where $\mathbf p^{\text{img}}_{i,k}$ is the crop’s centre in image co‑ordinates and $s_{i,k}$ is the scale factor (metres per pixel).  
      To limit the load on the subsequent vision–language model, we keep only the $N_{\max}=15$ largest candidates per frame.
\end{enumerate}

\noindent
The list of candidates is then mapped to the world frame through the same homography, and passed as lightweight, well‑structured data to the vision–language perception module described next.

\subsection{Vision–Language Model for Structured Vehicle Perception}\label{subsec:vlm_revised}

\begin{figure*}[ht]
    \centering
    \includegraphics[width=14cm]{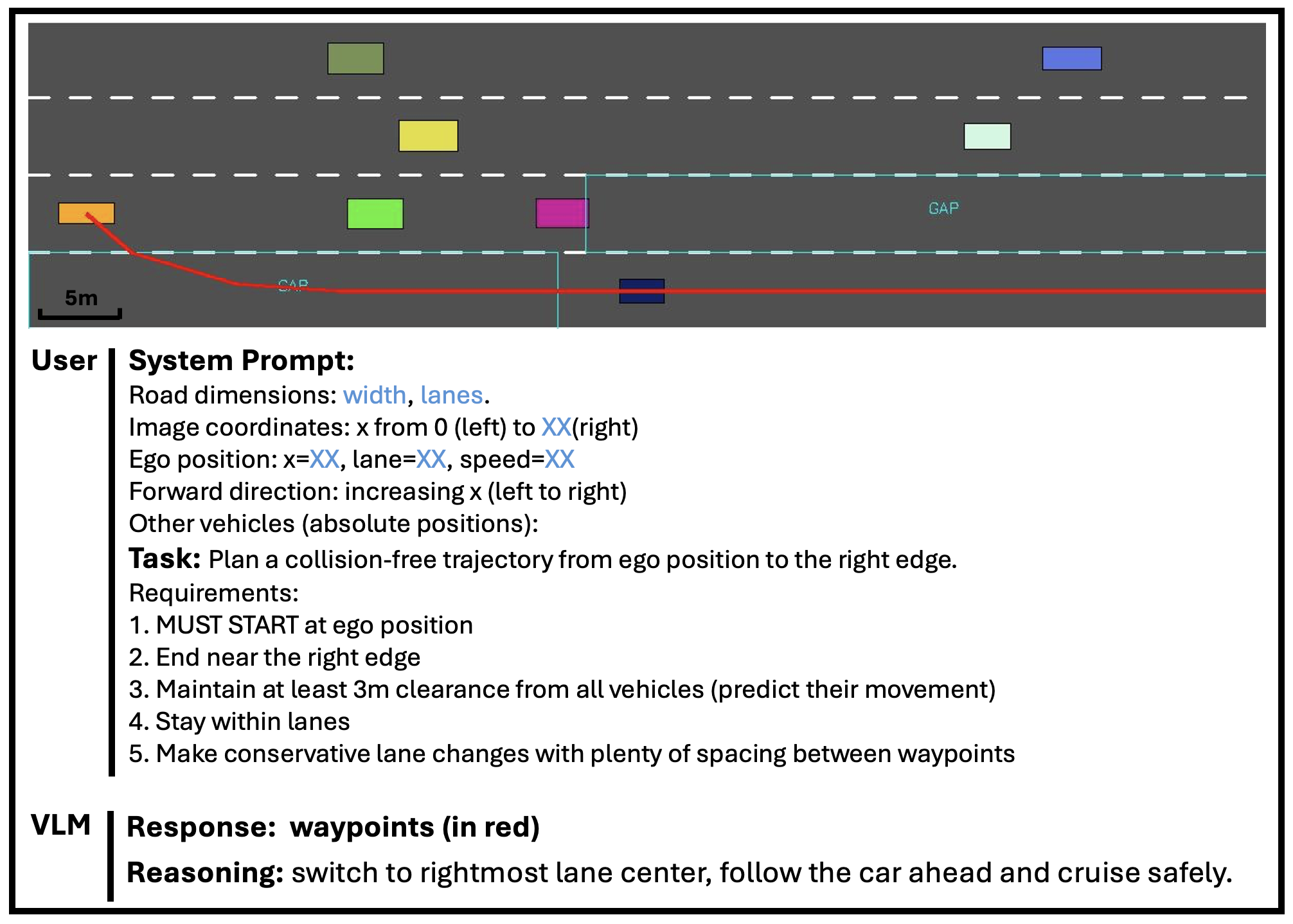}
    \caption{Vision-language models prompting.}
    \label{fig:VLM_prompt}
\end{figure*}

After the pre-processing module, each cropped patch of bounding boxes is forwarded together with its
world-frame offset to a commercial
VLM (OpenAI o4-mini-high \cite{openai_o4}) accessed through its public API.
We use the model in a \textbf{zero-shot} setting, i.e., without any
additional fine-tuning; the strong instruction-following capability of
the underlying architecture suffices to produce metrically consistent,
machine-readable descriptions of the vehicles depicted in each patch.
Every patch is accompanied by a template prompt, as shown in Fig. \ref{fig:VLM_prompt}.

\subsubsection{Two-stage decoding}
We prompt the model to follow a series of chain-of-thought steps \cite{wei2022chain}, 
\(\mathcal M_{\theta}\) first produces a short latent rationale  
(e.\,g., “\emph{single sedan oriented north-east}”),  
then outputs the structured JSON file.
If the output violates the key-value format,
an automatic re-prompt is issued until a valid message is obtained with a maximum iteration of 2.

For each valid entry, we form the output as the following vector
\[
   \mathbf o_{i,k}
   =\bigl[x^{o}_{i,k},\,y^{o}_{i,k},\,l_i,\,w_i\bigr]^{\!\top},
\]
where
\((x^{o}_{i,k},y^{o}_{i,k})\) is derived by adding the patch-local
offset to \(\mathbf p^{w}_{i,k}\) and scaling with \(s_{i,k}\).
A constant-velocity Kalman filter fuses two successive positions to
provide the velocity vector
\(
   \mathbf v_{i,k}
   =(\mathbf o^{p}_{i,k}-\mathbf o^{p}_{i,k-1})/T,
\)
with \(\mathbf o^{p}_{i,k}=[x^{o}_{i,k},y^{o}_{i,k}]^{\!\top}\).

\subsubsection{Interface to MPC}
The VLM module streams
\[
   \mathcal O_k
   =\bigl\{(\mathbf o_{i,k},\mathbf v_{i,k})\bigr\}_{i=1}^{N_k},
   \quad N_k\le N_{\max},~ N_k, N_{\max}\in\mathbb{N},
\]
to the optimization layer.
These structured states supply the geometric and kinematic information
required for the collision avoidance term used in the optimization problem as described in the following section.

\subsection{Collision Avoidance Potential Function}
\begin{figure}
    \centering
    \includegraphics[width=\linewidth]{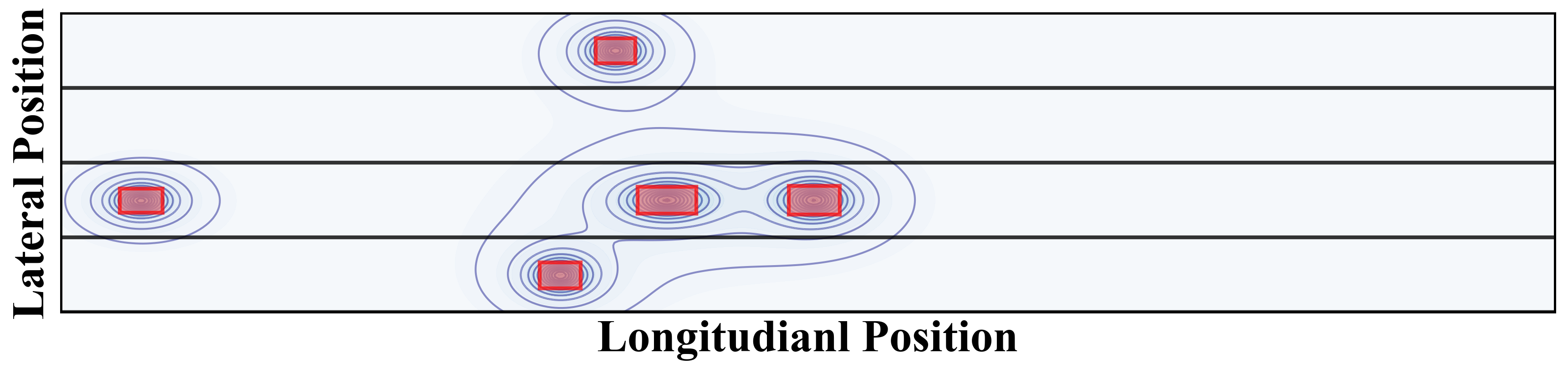}
    \caption{Example of the elliptical-shaped potential function.}
    \label{fig:pf}
\end{figure}

To represent obstacles in the optimization framework, we construct elliptical-shaped potential functions based on the information extracted by the VLM (see Fig. \ref{fig:pf}). For each detected obstacle vehicle (OV) $i$, we define a potential function $\Phi_i(\bm{x}_k)$ that establishes a safety zone around the vehicle:

\begin{equation}
\Phi_i(\bm{x}_k) = \exp\left[-\sqrt{\frac{(x_k - x_{i,k}^o)^2}{\sigma_{x,i}(\bm{x}_k)^2} + \frac{(y_k - y_{i,k}^o)^2}{\sigma_{y,i}^2}}\right],
\end{equation}
where $(x_{i,k}^o, y_{i,k}^o)$ represent the lateral and longitudinal positions of the OV $i$, and $\sigma_{x,i}$ and $\sigma_{y,i}$ define the dimensions of the ellipsoid. Parameter $\sigma_{y,i}$ defines the width of the ellipsoid and is set equal to the width of the lane, while $\sigma_{x,i}(\bm{x}_k)$ defines the length, which is calculated based on the time-gap policy as follows (see \cite{rajamani2011vehicle})

\begin{equation}
\sigma_{x,i}(\bm{x}_k) = \begin{cases}
    v_{x,k}\cdot \tau + L_i & \text{if \,} x_k \leq x_{i,k}^o, \\
    v_{x,i,k}^o\cdot \tau + L_i & \text{else},
\end{cases}
\end{equation}
where $\tau$ is a positive constant that represents the time gap.

\subsection{Optimization Problem}
\label{sec:optimization_problem}

The trajectory planning task is formulated as a finite-horizon discrete-time optimal control problem that integrates the collision avoidance potential functions derived from the VLM perception system. Building upon the system dynamics established in equations (1)-(2) and the control constraints defined in equations (3)-(6), the optimization seeks an optimal control sequence over the prediction horizon.

The optimization problem is defined as:

\begin{equation}
\begin{aligned}
\min_{\bm{u}} \quad & J = \sum_{k=0}^{K-1} L(\bm{x}_k, \bm{u}_k, k) \\
\text{s.t.} \quad & \bm{x}_{k+1} = \bm{f}(\bm{x}_k, \bm{u}_k), \quad k = 0, \ldots, K-1, \\
& \bm{x}_0 = \bm{x}(0), \\
& \bm{u}^{L}(\bm{x}_k) \leq \bm{u}_{k} \leq \bm{u}^{U}(\bm{x}_k), \quad k = 0, \ldots, K-1,
\end{aligned}
\end{equation}
where $K\in\mathbb{N}$ represents the prediction horizon, and the system dynamics and control bounds follow the formulations presented in Section \ref{sec:problem_formulation}.

The stage cost function balances multiple objectives including reference tracking, control smoothness, speed regulation, and obstacle avoidance:

\begin{equation}
\begin{aligned}  \label{eq:obj_func}
L(\bm{x}_k,& \bm{u}_k, k) = p_1 u_{x,k}^2 + p_2 u_{y,k}^2 + p_3 (v_{x,k} - v_{des})^2 \\
&+ p_4 v_{y,k}^2 + \sum_{i=1}^{N_{obs}} \lambda_i \Phi_i(\bm{x}_k),
\end{aligned}
\end{equation}
where the cost weights $p_1, p_2, p_3,$ and $ p_4$ prioritize the minimization of the control effort, reflecting fuel efficiency \cite{typaldos2020minimization}; speed tracking; and lateral stability, respectively. The collision avoidance potential functions $\Phi_i(\bm{x}_k)$, defined in (8), provide smooth repulsive forces that guide the optimizer away from detected obstacles. The weighting factors $\lambda_i > 0$ determine the influence of each obstacle on the cost function.

This formulation differs from traditional MPC approaches by incorporating vision-derived obstacle information directly into the cost function through the potential functions $\Phi_i(\bm{x}_k)$. The adaptive elliptical parameterization described in equation (9) ensures that safety zones dynamically adjust based on relative motion between the ego vehicle and obstacles, providing enhanced protection in high-speed scenarios while maintaining computational efficiency.

The resulting optimization problem is nonlinear due to the exponential form of the potential functions and the state-dependent control bounds. The problem is solved using the constrained DDP algorithm, as detailed in Section \ref{sec:ddp_solution}, where the potential functions are integrated as soft constraints within the optimization framework.

\subsection{Differential Dynamic Programming}
\label{sec:ddp_solution}

For the solution of the optimization problem defined in Section \ref{sec:optimization_problem}, our approach employs the DDP algorithm. DDP is a second-order optimization method for solving nonlinear optimal control problems without requiring discretization of the state and control spaces. In this paper, we employ an extension of DDP that accounts for inequality constraints.

Consider a discrete-time optimal control problem with states $\bm{x}$ and controls $\bm{u}$ with the recursive Bellman equation 
\begin{align}
    V_k(\bm{x}_k)&=\min_{\bm{u}_k} Q_k(\bm{x}_k, \bm{u}_k), \\
    Q_k(\bm{x}_k, \bm{u}_k)&=L(\bm{x}_k,\bm{u}_k) + V_{k+1}(\bm{f}(\bm{x}_k,\bm{u}_k)) \label{eq: bellman},  
\end{align}
where $V_k$ is the optimal cost function for $k=K-1,\dots,0$, $L(\cdot)$ is the objective function of the OCP \eqref{eq:obj_func}, $\bm{f}$ is the right-hand side of the state equations \eqref{eq:states_gen} and \eqref{eq:states}, and the minimization must be carried out for all feasible controls $\bm{u}_k$ that satisfy any inequality constraints present. The DDP procedure performs at each time step of each iteration, a quadratic approximation of the term to be minimized in the recursive Bellman equation \eqref{eq: bellman}. The quadratic approximation is taken around nominal trajectories ($\bm{\bar{x}}_k, \bm{\bar{u}}_k$), which are the initial trajectories of each iteration. Define $\delta \bm{x}_k = \bm{x}_k- \bm{\bar{x}}_k$ and $\delta \bm{u}_k = \bm{u}_k - \bm{\bar{u}}_k$. The aim is to find the optimal control law for $\delta \bm{{u}}_k$, which minimizes the quadratic approximation, subject to some inequality constraints. The procedure of the DDP algorithm for each iteration is described in brief as follows (see \cite{yakowitz1986stagewise, murray1979constrained} for more details). 

\subsubsection{Backward Pass}
During the \textit{backward} pass the quadratic approximation of $Q_k(\bm{x}_k,\bm{u}_k)$ around a nominal point ($\bm{\bar{x}}_k,\bm{\bar{u}}_k$), is given by
\begin{equation}
\begin{aligned}
    Q_k(\delta&\bm{x}_k,\delta\bm{u}_k) \approx\ 
Q^T_{\bm{x},k} \delta\bm{x}_k + Q^T_{\bm{u},k} \delta\bm{u}_k
+ \dfrac{1}{2} \delta\bm{x}_k^T Q_{\bm{x}\bm{x},k} \delta\bm{x}_k \\
&+ \delta\bm{x}_k Q_{\bm{x}\bm{u},k} \delta\bm{u}_k + \dfrac{1}{2} \delta\bm{u}^T_k Q_{\bm{u}\bm{u},k} \delta\bm{u}_k,
\end{aligned}
\end{equation}
where $Q_{\cdot,k}$ are the coefficient matrices, derived as follows
\begin{align}
Q_{x,k} &= L_{\bm{x}} + \bm{f}_x^T V_{x,k+1}, \\
Q_{u,k} &= L_{\bm{u}} + \bm{f}_u^T V_{x,k+1},\\
Q_{xx,k} &= L_{xx} + \bm{f}_x^T V_{xx,k+1} \bm{f}_x + V_{x,k+1} \bm{f}_{xx},\\
Q_{ux,k} &= L_{ux} + \bm{f}_u^T V_{xx,k+1} \bm{f}_x + V_{x,k+1} \bm{f}_{ux},\\
Q_{uu,k} &= L_{uu} + \bm{f}_u^T V_{xx,k+1} \bm{f}_u + V_{x,k+1} \bm{f}_{uu},
\end{align}
where $V_{xx}$ and $V_{x}$ are the Hessian matrix and the gradient vector, respectively. The derivatives of the cost and state equations are computed at the nominal point $(\bm{\bar{x}}_k, \bm{\bar{u}}_k)$.

The unconstrained optimal control perturbation is then given by
\begin{equation}
\delta\bm{u}^*_k = \bm{k}_k + \bm{K}_k\delta\bm{x}_k,
\end{equation}
where
\begin{equation}
\begin{aligned}
    \bm{k}_k &= -Q_{uu,k}^{-1}Q_{u,k}, \\ 
    \bm{K}_k &= -Q_{uu,k}^{-1}Q_{ux,k}.
\end{aligned}
\end{equation}
This minimization outcome is substituted in the quadratic approximation $Q$ to obtain, from the Bellman equation, the approximate optimal value function
\begin{equation}
V_k\left(\bm{x}(k)\right) = Q_k\left(\bm{x}_k,\bm{k}_k+\bm{K}_k\delta\bm{x}_k\right).	
\end{equation}

\subsubsection{Forward Pass}
The nominal trajectories are updated as
\begin{align}
&\bm{u}_k = \bar{\bm{u}}_k + \varepsilon\Bigl[\bm{k}_k+\bm{K}_k\,\delta\bm{x}(k)\Bigr], \label{eq:forward1}\\
&\bm{x}_{k+1} = \bm{f}\Bigl(\bm{x}_k,\bm{u}_k\Bigr), \label{eq:forward2}\\
&\bm{x}_0: \text{given initial states}, \label{eq:forward3}
\end{align}
with a step-size \(0<\epsilon\le1\) chosen via line search. Convergence is declared when
\begin{equation} \label{eq:convergence}
\left(\sum_{k=0}^{K-1}\|\bm{u}(k)-\bar{\bm{u}}(k)\|^2\right)^{1/2} < \varepsilon_1,
\end{equation}
with \(\varepsilon_1>0\) a small threshold.

\subsubsection{Constrained DDP}
The standard DDP method \cite{mayne1970} does not account for constraints; thus, we extend DDP using nonlinear constraints  \cite{yakowitz1986stagewise, xie2017differential}.

The control constraints are given by
\[
\bm{c}\bigl(\bm{x}_k,\bm{u}_k\bigr) \le \bm{0}.
\]
Let \(\widetilde{\bm{c}}(\bm{x},\bm{u})\) be the linearization of \(\bm{c}(\bm{x},\bm{u})\) about \((\bar{\bm{x}}_k,\bar{\bm{u}}_k)\). The quadratic programming subproblem in the backward pass is
\begin{equation}
\begin{aligned}
\min_{\delta\bm{u}_k} \quad & Q\Bigl(\bar{\bm{x}}_k,\bar{\bm{u}}_k+\delta\bm{u}_k\Bigr),\\[1mm]
\text{s.t.} \quad & \widetilde{\bm{c}}\Bigl(\bar{\bm{x}}_k,\bar{\bm{u}}_k+\delta\bm{u}_k\Bigr) \leq \bm{0}.
\end{aligned}
\end{equation}
Let \(S\) denote the active set of constraints (i.e., those with \(\widetilde{c}_i = 0\)); these can be written as
\begin{equation} \label{eq:activeSet}
\bm{U}\,\delta\bm{u}_k - \bm{W} = \bm{0}.
\end{equation}
The necessary conditions for optimality are derived from the Lagrangian
\begin{equation} \label{eq:lagrangian}
\begin{aligned}
    \mathcal{L}\bigl(\delta\bm{u}_k,\bm{\lambda}_k\bigr) &= \frac{1}{2}\delta\bm{u}_k^T\bm{Q}_{\bm{uu},k}\delta\bm{u}_k + \bm{Q}_{\bm{u},k}^T\delta\bm{u}_k \\
    &+ \bm{\lambda}_k^T\Bigl(\bm{U}\,\delta\bm{u}_k-\bm{W}\Bigr).
\end{aligned}
\end{equation}
Differentiating \eqref{eq:lagrangian} with respect to \(\delta\bm{u}_k\) and \(\bm{\lambda}_k\) yields
\begin{equation} \label{eq:langSystem}
\begin{bmatrix}
\bm{Q}_{\bm{uu},k} & \bm{U}\\
\bm{U}^T & \bm{0}
\end{bmatrix}
\begin{bmatrix}
\delta\bm{u}_k\\
\bm{\lambda}_k
\end{bmatrix}
=
\begin{bmatrix}
-\bm{Q}_{\bm{u},k}\\
\bm{W}
\end{bmatrix}.
\end{equation}
If we assume that the active constraints remain valid for any \(\bm{x}_k=\bar{\bm{x}}_k+\delta\bm{x}_k\), then
\begin{equation}
\bm{U}\,\delta\bm{u}_k-\bm{W}-\bm{X}\,\delta\bm{x}_k=\bm{0},
\end{equation}
which leads to an affine control law
\begin{equation} \label{eq:controlLaw}
\delta\bm{u}_k=\bm{k}_k+\bm{K}_k\,\delta\bm{x}_k.
\end{equation}

\subsubsection{Regularization}
To ensure numerical stability and convergence of the DDP algorithm, particularly when the Hessian matrices $Q_{\bm{uu},k}$ become ill-conditioned or singular, a regularization scheme is employed. The regularized Hessian is given by:
\begin{equation}
Q_{\bm{uu},k}^{\text{reg}} = Q_{\bm{uu},k} + \mu \bm{I},
\end{equation}
where $\mu > 0$ is the regularization parameter and $\bm{I}$ is the identity matrix.
The regularization parameter is adaptively adjusted during the iteration process according to the following strategy in Algorithm \ref{alg:reg}. 

\begin{algorithm}
\caption{Regularization procedure}
\label{alg:reg}
\begin{algorithmic}
\STATE \textbf{Initialization:} Set $\mu = \mu_{\min} = 10^{-6}$
\STATE \textbf{Backward pass failure:} If the backward pass fails due to numerical issues (e.g., matrix singularity, negative definite Hessian, or constraint solving failure), increase the regularization: $\mu \leftarrow \min(\mu \cdot \gamma, \mu_{\max})$ where $\gamma = 5$ is the regularization increase factor and $\mu_{\max} = 10^6$
\STATE \textbf{Forward pass success:} If the forward pass successfully improves the cost, decrease the regularization: $\mu \leftarrow \max(\mu / \gamma, \mu_{\min})$
\STATE \textbf{Algorithm termination:} If $\mu \geq \mu_{\max}$, the algorithm terminates as the problem is considered too ill-conditioned to solve reliably
\end{algorithmic}
\end{algorithm}

The adaptive regularization strategy dynamically adjusts the regularization parameter $\mu$ based on the algorithm's performance at each iteration. Initially, $\mu$ is set to a small value ($\mu_{\min} = 10^{-6}$) to minimize interference with the optimization when the Hessian is well-conditioned. When the backward pass encounters numerical difficulties, e.g., matrix singularity, a negative definite Hessian, or constraint-solving failures, the algorithm interprets this as evidence that the current Hessian $Q_{uu,k}$ is ill-conditioned. In response, it increases the regularization parameter by multiplying it by a factor $\gamma = 5$, up to a maximum value $\mu_{\max} = 10^6$. This increase in $\mu$ adds more weight to the identity matrix in (32), improving the condition number of the regularized Hessian $Q^{reg}_{uu,k}$ and restoring numerical stability. In contrast, when the forward pass successfully reduces the cost function, indicating that the problem is well-conditioned and the algorithm is making progress, $\mu$ is decreased by dividing it by $\gamma$, allowing the algorithm to rely more heavily on the original Hessian information. If $\mu$ reaches the maximum threshold $\mu_{\max}$, the algorithm terminates, assuming that the problem has become too ill-conditioned to solve reliably, thereby preventing infinite loops and computational waste on intractable problems. This adaptive regularization ensures that the algorithm remains numerically stable while maintaining computational efficiency when the problem is well-conditioned.

\subsubsection{DDP Pseudocode and procedure}

The complete DDP algorithm with adaptive regularization for solving the constrained optimal control problem is summarized in Algorithm \ref{alg:DDP}.

\begin{algorithm}[h!]
\caption{DDP algorithmic steps}
\label{alg:DDP}
\begin{algorithmic}
\REQUIRE Initial state $\bm{x}_0$, initial control sequence $\{\bm{u}_k^{(0)}\}_{k=0}^{K-1}$, tolerance $\varepsilon_1$
\ENSURE Optimal control sequence $\{\bm{u}_k^{*}\}_{k=0}^{K-1}$ and state trajectory $\{\bm{x}_k^{*}\}_{k=0}^{K}$
\STATE Initialize: $\mu = \mu_{\min} = 10^{-6}$, $n = 0$
\STATE Compute initial trajectory from initial control sequence
\WHILE{not converged and $n < N_{\max}$}
\WHILE{\textit{backward\_pass\_failed} and $\mu < \mu_{\max}$}
\STATE \textbf{Backward Pass:}
\FOR{$k = K-1, \ldots, 0$}
\STATE Compute derivatives: $Q_{\bm{x},k}$, $Q_{\bm{u},k}$, $Q_{\bm{xx},k}$, $Q_{\bm{ux},k}$, $Q_{\bm{uu},k}$
\STATE Apply regularization: $Q_{\bm{uu},k} \leftarrow Q_{\bm{uu},k} + \mu \bm{I}$
\STATE Solve constrained QP for $\delta\bm{u}_k$ with box constraints
\STATE Compute feedback gains: $\bm{k}_k$, $\bm{K}_k$
\ENDFOR
\IF{backward pass successful}
\STATE \textit{backward\_pass\_failed} $\leftarrow$ \textbf{false}
\ELSE
\STATE $\mu \leftarrow \min(5\mu, 10^6)$ \COMMENT{Increase regularization}
\ENDIF
\ENDWHILE

\STATE \textbf{Forward Pass:}
\FOR{each step size $\alpha \in \{1.0, 0.5, 0.1, 0.05, 0.01\}$}
    \STATE Update controls: $\bm{u}_k^{\text{new}} = \bm{u}_k + \alpha(\bm{k}_k + \bm{K}_k \delta\bm{x}_k)$
    \STATE Simulate trajectory and compute cost $J^{\text{new}}$
    \IF{$J^{\text{new}} < J^{(n)}$}
        \STATE Accept new trajectory: $\bm{u}_k^{(n+1)} = \bm{u}_k^{\text{new}}$
        \STATE $\mu \leftarrow \max(\mu/5, 10^{-6})$ \COMMENT{Decrease regularization}
        \STATE \textbf{break}
    \ENDIF
\ENDFOR

\STATE Check convergence: $\left\|\bm{u}^{(n+1)} - \bm{u}^{(n)}\right\| < \varepsilon_1$
\STATE $n \leftarrow n + 1$
\ENDWHILE
\STATE \textbf{return} $\{\bm{u}_k^{(n)}\}_{k=0}^{K-1}$, $\{\bm{x}_k^{(n)}\}_{k=0}^{K}$
\end{algorithmic}
\end{algorithm}

At the start of each iteration, the algorithm first retrieves the latest structured scene and vehicle state information from VLM, using it to seed the initial trajectory guess and update the collision-avoidance constraints. The algorithm begins with an initial guess for the control sequence and iteratively refines it through backward and forward passes until convergence is achieved. The backward pass computes the quadratic approximation of the $ Q$ function at each time step by calculating the necessary derivatives of the cost function and the system dynamics. The regularization term is applied to ensure numerical stability when the Hessian matrix $Q_{\bm{uu},k}$ becomes ill-conditioned. For each time step, the algorithm solves a constrained quadratic programming subproblem to determine the optimal control perturbation $\delta\bm{u}_k$ while respecting the box constraints derived from the state-dependent control bounds. The forward pass performs a line search to update the nominal trajectory. Multiple-step sizes are tested to ensure sufficient cost reduction while maintaining feasibility. If the forward pass succeeds in reducing the cost, the regularization parameter is decreased to encourage faster convergence. Conversely, if no improvement is found, regularization is increased to enhance numerical stability. The algorithm terminates when either the convergence criterion is satisfied (the norm of control changes falls below the threshold $\varepsilon_1$). 
This adaptive regularization scheme is crucial to handle the non-linear nature of obstacle avoidance constraints and ensures robust performance in various traffic scenarios.

\subsection{Safety Verification}
To ensure the reliability of the obtained optimal trajectories from the DDP algorithm, we implement a safety verification layer that evaluates the obtained trajectories over multiple safety criteria. This verification mechanism provides a quick assessment of potential collisions or safety violations over the finite time horizon, serving as a critical safeguard before the trajectory execution.

Let $\{\bm{x}^*_{k}\}_{k=0}^{K-1}$ denote the optimal state trajectory obtained from the DDP solution, where $\bm{x}^*_{k} = [x_k^*, y_k^*, v_{x,k}^*, v_{y,k}^*]^T$ represents the ego vehicle's state at discrete time step $k$. The safety verification evaluates this trajectory against the predicted motion of surrounding vehicles within a verification horizon $T_v \leq T_h$, where $T_h$ is the MPC prediction horizon.

For each detected obstacle vehicle $j \in \{1, \ldots, N_{obs}\}$, the current state is characterized by position $\bm{p}_{j,0} = [x_{j,0}, y_{j,0}]^T$, velocity $\bm{v}_{j,0} = [v_{x,j,0}, v_{y,j,0}]^T$, and dimensions (length $L_j$ and width $W_j$). The ego vehicle dimensions are denoted by $L_{ego}$ and $W_{ego}$.

The predicted position of obstacle vehicle $j$ at discrete time step $m$ within the verification horizon is computed using a constant velocity model:
\begin{equation}
\bm{p}_{j,m} = \bm{p}_{j,0} + m \cdot T \cdot \bm{v}_{j,0}, \quad m = 1, \ldots, M,
\end{equation}
where $M = \lfloor T_v / T \rfloor$ and $T$ is the sampling period.

For collision detection, we model each vehicle as an oriented bounding box. The bounding box for the ego vehicle at time step $m$ is defined as:
\begin{equation}
\begin{aligned}
\mathcal{B}_{ego}(m) = \Bigg\{\bm{p} \in \mathbb{R}^2 :& \left|p_x - x_m^*\right| \leq \frac{L_{ego}}{2}, \\
 & \left|p_y - y_m^*\right| \leq \frac{W_{ego}}{2}\Bigg\},
\end{aligned}
\end{equation}
and similarly for obstacle vehicle $j$:
\begin{equation}
\begin{aligned}
\mathcal{B}_{j}(m) = \Bigg\{\bm{p} \in \mathbb{R}^2 :& \left|p_x - x_{j,m}\right| \leq \frac{L_j}{2},  
    \\ & \left|p_y - y_{j,m}\right| \leq \frac{W_j}{2}\Bigg\}.
\end{aligned}
\end{equation}

The safety verification evaluates four critical criteria at each discrete time step $m \in \{1, \ldots, M\}, M\in\mathbb{N}$. First, direct collision is detected when bounding boxes intersect, as defined by the indicator function:
\begin{equation}
\mathcal{C}_j(m) = \begin{cases}
1 & \text{if } \mathcal{B}_{ego}(m) \cap \mathcal{B}_j(m) \neq \emptyset, \\
0 & \text{otherwise}.
\end{cases}
\end{equation}

Second, longitudinal safety is assessed through time-to-collision (TTC) analysis. For obstacles ahead of the ego vehicle (i.e., $x_{j,m} > x_m^*$), we compute
\begin{equation}
d_{lon,j}(m) = (x_{j,m} - \frac{L_j}{2}) - (x_m^* + \frac{L_{ego}}{2}),
\end{equation}
\begin{equation}
TTC_j(m) = \begin{cases}
\frac{d_{lon,j}(m)}{v_{x,m}^* - v_{x,j,0}} & \text{if } v_{x,m}^* > v_{x,j,0}, \\
\infty & \text{otherwise}.
\end{cases}
\end{equation}
The longitudinal safety criterion is violated when $TTC_j(m) < TTC_{min}$:
\begin{equation}
\mathcal{L}_j(m) = \begin{cases}
1 & \text{if } TTC_j(m) < TTC_{min}, \\
0 & \text{otherwise}.
\end{cases}
\end{equation}

Third, lateral safety examines the clearance between vehicles. The lateral separation is computed by
\begin{equation}
d_{lat,j}(m) = |y_m^* - y_{j,m}| - \frac{W_{ego} + W_j}{2},
\end{equation}
and the safety criterion is violated when clearance is insufficient by
\begin{equation}
\mathcal{S}_j(m) = \begin{cases}
1 & \text{if } d_{lat,j}(m) < d_{lat,min}, \\
0 & \text{otherwise}.
\end{cases}
\end{equation}

Fourth, road boundary compliance ensures the ego vehicle remains within designated boundaries $\mathcal{R}_{road}$, i.e.,
\begin{equation}
\mathcal{R}(m) = \begin{cases}
1 & \text{if } \mathcal{B}_{ego}(m) \not\subseteq \mathcal{R}_{road}, \\
0 & \text{otherwise}.
\end{cases}
\end{equation}

The overall safety assessment aggregates all criteria over the verification horizon. The trajectory is classified as unsafe if any collision or road boundary violation occurs:
\begin{equation}
\text{Unsafe} = \bigvee_{m=1}^{M} \left[\mathcal{R}(m) \vee \bigvee_{j=1}^{N_{obs}} \mathcal{C}_j(m)\right],
\end{equation}
and as high-risk if safety margins are insufficient:
\begin{equation}
\text{High-Risk} = \bigvee_{m=1}^{M} \bigvee_{j=1}^{N_{obs}} [\mathcal{L}_j(m) \vee \mathcal{S}_j(m)].
\end{equation}

When either condition is satisfied (Unsafe $\vee$ High-Risk), the verification layer triggers a replanning request to the MPC framework with modified constraints or cost function parameters. The safety verification operates with a verification horizon of $T_v = 3.0$ s, minimum TTC threshold of $TTC_{min} = 2.0$ s, minimum lateral clearance of $d_{lat,min} = 0.5$ m, and a sampling period of $T = 0.1$ s. While this verification layer provides a quick safety assessment rather than formal guarantees, it serves as a safeguard that enhances the overall robustness of the proposed approach by detecting and addressing potential collisions and safety violations.

\subsection{MPC Framework}
\label{sec:mpc_framework}

The MPC framework serves as the central coordination mechanism that integrates VLMs, optimal control, and safety verification into a unified real-time trajectory planning system. Operating on a receding horizon principle, the framework employs an event-triggered replanning strategy to achieve computational efficiency while maintaining safety and performance guarantees in dynamic traffic environments. 

\begin{figure}
    \centering
    \includegraphics[width=1\linewidth]{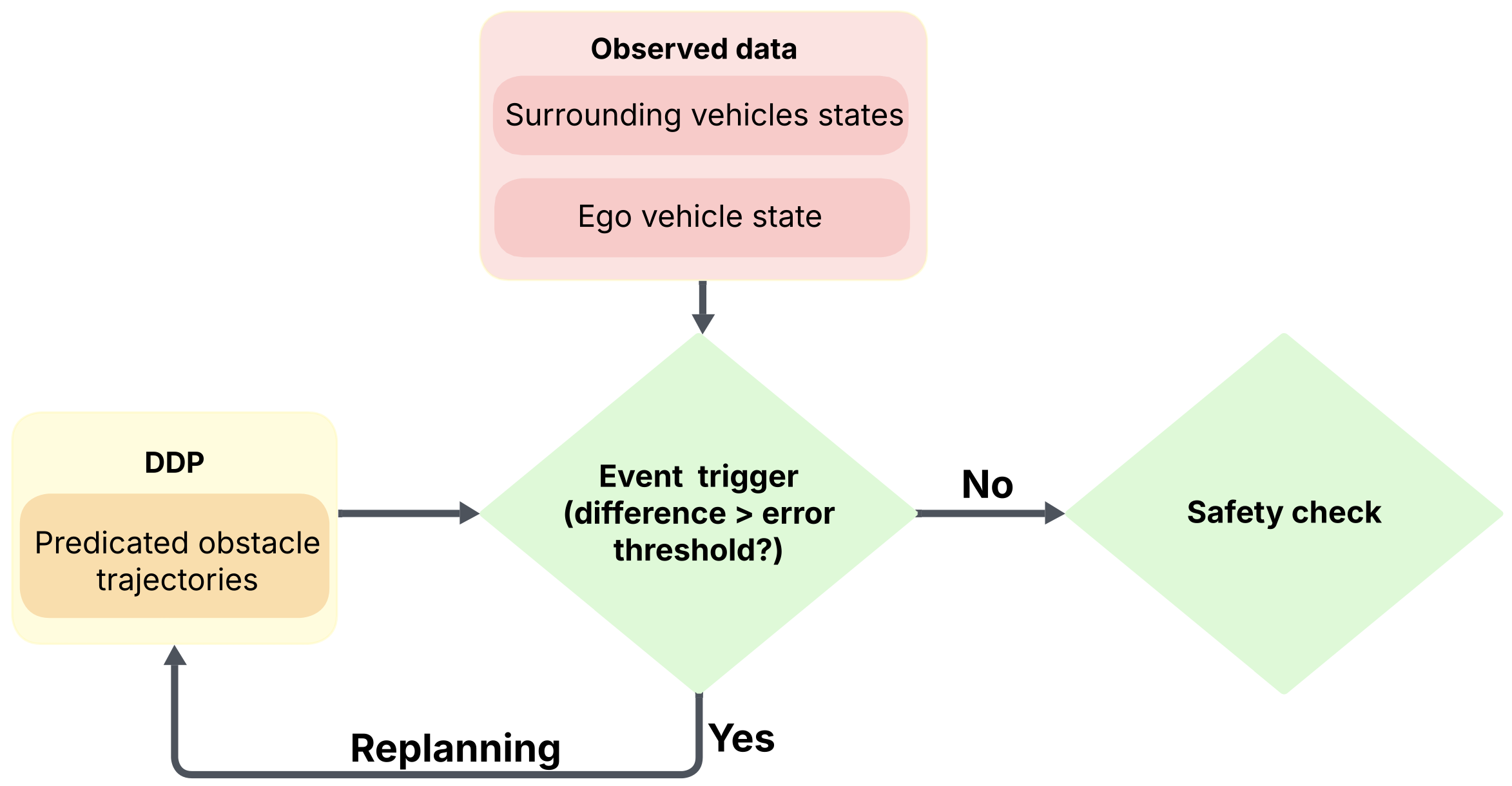}
    \caption{Framework of MPC (event-trigger based).}
    \label{fig:mpc_framwork}
\end{figure}

The framework departs from conventional fixed-interval MPC approaches by implementing an adaptive replanning mechanism that responds to environmental changes rather than operating on predetermined time intervals. This event-driven approach significantly reduces computational burden while ensuring responsive adaptation to dynamic traffic scenarios. As illustrated in Fig. \ref{fig:mpc_framwork}, the framework continuously monitors the environment and makes replanning decisions based on predefined criteria, executing optimization only when necessary.

When environmental changes exceed predefined thresholds, the system proceeds to replan using DDP with predicted obstacle trajectories. Otherwise, the framework continues with safety verification of the current trajectory, ensuring efficient computational resource utilization while maintaining safety guarantees.

The event-triggered replanning strategy evaluates multiple environmental and vehicle state conditions to determine when trajectory recomputation is warranted. Replanning is initiated when any of the following conditions are satisfied:

\begin{equation}
\text{Replan}(t_k) = \bigvee_{i=1}^{4} C_i(t_k).
\end{equation}

The individual criteria encompass temporal, spatial, and behavioral triggers:
\begin{align}
C_1(t_k) &: \quad t_k - t_{\text{last}} \geq T_h, \\
C_2(t_k) &: \quad |\mathcal{O}_k| > |\mathcal{O}_{k-1}|, \\
C_3(t_k) &: \quad \max_{i} \|\bm{p}_{i,k} - \hat{\bm{p}}_{i,k}\| > 2.0 \text{ m}, \\
C_4(t_k) &: \quad \max_{i} |l_{i,k} - l_{i,k-1}| \geq 1,
\end{align}
where $T_h$ represents the maximum planning horizon, $\mathcal{O}_k$ denotes the set of detected obstacles at time $t_k$, $\bm{p}_{i,k}$ and $\hat{\bm{p}}_{i,k}$ are the actual and predicted positions of obstacle $i$, and $l_{i,k}$ represents the lane index.

The first criterion $C_1$ ensures that replanning occurs when the planning horizon has elapsed, guaranteeing that the trajectory remains valid within the predictive time window. Criterion $C_2$ triggers replanning when new obstacles enter the detection zone, requiring immediate trajectory adjustment to account for previously unknown threats. The third criterion $C_3$ monitors the deviation between predicted and actual obstacle positions, initiating replanning when obstacles deviate significantly from their expected trajectories, indicating potential changes in driving behavior or vehicle dynamics. Finally, $C_4$ detects lane changes by neighboring vehicles, which fundamentally alter the traffic configuration and require immediate consideration in the trajectory planning of the ego vehicle.

To prevent computational instability from frequent replanning due to measurement noise or minor disturbances, a minimum replanning interval constraint $\Delta t_{\min} = 1.0$ s is enforced.

When replanning is triggered, the framework executes a structured optimization cycle that begins with a comprehensive assessment of the state and environment. The states of the ego vehicle $\bm{x}_k = [x_k, y_k, v_{x,k}, v_{y,k}]^T$ are obtained directly from the simulation environment, while the obstacle set $\mathcal{O}_k$ is extracted from the VLM perception system. This dual-source approach ensures accurate state estimation while leveraging the advanced scene understanding capabilities of vision-language models.

Given the computational complexity of handling numerous obstacles in dense traffic scenarios, the framework implements a strategic obstacle prioritization scheme. This prioritization focuses computational resources on the most critical threats while maintaining tractable problem dimensions:

\begin{equation}
\text{Priority}(o_i) = \begin{cases}
x_{i,k} - x_{\text{ego},k} & \text{if } x_{i,k} \geq x_{\text{ego},k}, \\
x_{i,k} - x_{\text{ego},k} - 1000 & \text{if } x_{i,k} < x_{\text{ego},k}.
\end{cases}
\end{equation}

This formulation ensures that vehicles ahead of the ego vehicle, which poses greater collision risks, receive higher priority in the optimization process. The large negative offset for vehicles behind effectively deprioritizes them while maintaining them in the obstacle set for safety considerations.

The trajectory initialization process takes advantage of the available VLM-generated waypoints when possible to improve the convergence characteristics and the quality of the solution. When VLM trajectory data are available, the framework applies spline-based smoothing to generate a kinematically feasible reference trajectory, subsequently converting this trajectory to control inputs through numerical differentiation:

\begin{equation}
\bm{u}_{\text{init},k} = \frac{\bm{v}_{k+1} - \bm{v}_k}{T},
\end{equation}
where $\bm{v}_k$ represents the velocity vector at time step $k$ and $T$ is the discretization time interval. In some scenarios where VLM data is unavailable or unreliable, for example, if the initial trajectory generated by VLM collides with surrounding vehicles, the framework defaults to zero initialization: $\bm{u}_{\text{init}} = \bm{0}$, allowing the optimizer to determine the trajectory from the current state.

Following trajectory optimization through the DDP algorithm described in previous sections, the framework applies the first control input according to the receding horizon principle. Speed commands incorporate smoothing to prevent abrupt velocity changes that could affect passenger comfort or vehicle stability:

\begin{equation}
v_{\text{command}} = v_{\text{current}} + \text{sat}[v_{\text{target}} - v_{\text{current}}; -1, 1],
\end{equation}
where $\text{sat}[\ \cdot \ ; a, b]$ denotes the saturation function that limits the argument between bounds $a$ and $b$. This formulation restricts speed changes to 1.0 m/s per time step, ensuring smooth acceleration and deceleration profiles.

The framework incorporates adaptive speed management mechanisms to respond intelligently to varying traffic conditions. The leader-following behavior activates when a vehicle is detected within a 50-meter range ahead:

\begin{figure*}[ht]
    \centering
    \includegraphics[width=1\textwidth]{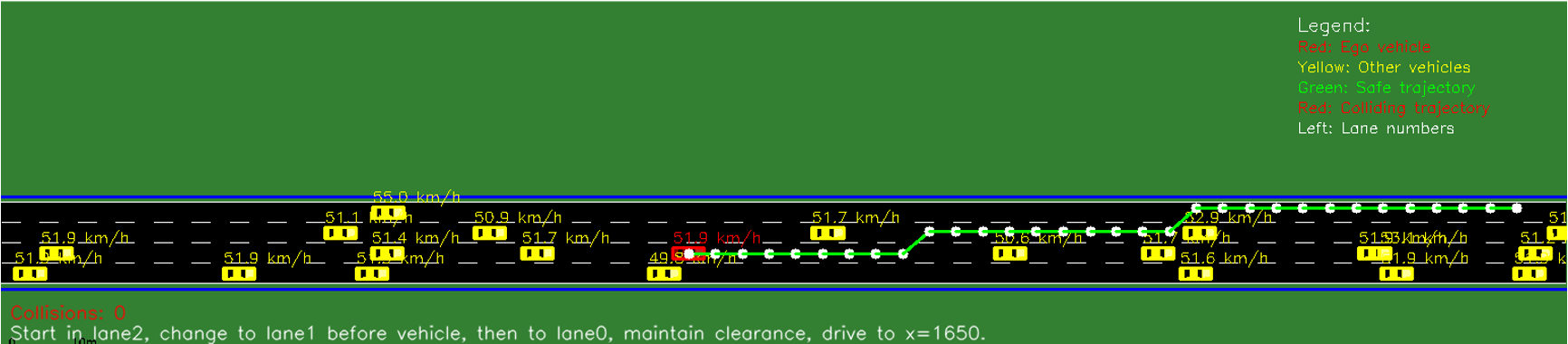}
    \caption{VLM initial trajectory generation.}
    \label{fig:vlm_out}
\end{figure*}

\begin{equation}
v_{\text{desired}} = \begin{cases}
\min(v_{\text{nominal}}, 0.95 v_{\text{leader}}), & \text{if } d \leq 20 \text{ m}, \\
\min(v_{\text{nominal}}, w \cdot 0.95 v_{\text{leader}}, & \text{if } 20 < d \leq 50 \text{ m}, \\
    \qquad + (1-w) v_{\text{current}}),
\end{cases}
\end{equation}
where the blending weight $w = 1 - \frac{d - 20}{30}$ provides smooth transitions between free-flow and following behaviors. Additionally, progressive acceleration simulates realistic driving behavior by incrementally increasing the desired speed every 4 seconds by 0.5 m/s until reaching the target velocity.

To maintain real-time performance, the framework employs several computational optimization strategies. Warm starting utilizes previous optimal trajectories as initial guesses for subsequent optimizations, significantly reducing convergence time. Selective obstacle processing limits the number of tracked vehicles based on their criticality scores, preventing computational explosion in dense traffic scenarios while maintaining safety.

This event-triggered MPC framework successfully demonstrates the integration of modern AI perception capabilities with traditional optimal control methods, achieving a better improvement in mixed-traffic efficiency compared to fixed-interval replanning while maintaining equivalent safety and performance standards. The framework's ability to adapt to dynamic traffic conditions while ensuring real-time operation represents a significant advancement in autonomous vehicle control systems for mixed traffic environments.

\section{EXPERIMENT and RESULTS}
This section details the experimental setup, the implemented trajectory planning framework, simulation scenarios, and the evaluation methodology used to assess the performance of our proposed system. All experiments were conducted using the Simulation of Urban Mobility (SUMO) environment, with its Python API, Traffic Control Interface (Traci), for dynamic control and data acquisition.

\subsection{Environment Setup}
We conducted all simulation experiments on a four-lane freeway with a horizon of 1000 seconds. The first 500 seconds served as a warm-up period to ensure more precise image processing, followed by 7 minutes of active simulation. The warm-up period is needed to ensure that experimental results reflect the system's true steady-state capabilities rather than the potentially inconsistent performance during the initial startup and calibration phases. We designated each ego car to travel through a 2 km segment before exiting the simulation. Upon exit, a new ego car was selected. This approach allowed us to generate diverse scenarios to validate our proposed framework thoroughly, with traffic compositions including different types of cars and trucks at varying demand levels as detailed in Table \ref{tab:traffic_flow}. The experiments utilized different flow densities to evaluate the robustness of our proposed framework. Additionally, we compared our method against several alternatives: a baseline approach, the baseline approach with image processing augmentation, and the baseline approach with VLM. Our framework integrates vision-based initial planning with a DDP controller managed by an MPC replanning scheme.

\subsection{Vision-Based Initial Trajectory Generation}

An initial trajectory can be generated using a VLM, completed through the following steps: First, we capture a screenshot of the SUMO GUI to obtain a BEV map. We then process the image to detect road boundaries, the number of lanes with corresponding widths, and the bounding boxes of surrounding vehicles. Vehicle speeds are estimated based on position history. These steps are performed on an RTX A6000 GPU in $<20$\, ms. Next, we construct a prompt providing road dimensions, ego state, and other vehicle information. The VLM returns an initial trajectory with a reasoning interpretation for the DDP controller as shown in Fig. \ref{fig:vlm_out}. This process aims to provide a feasible starting point for the optimization, potentially reducing convergence time and improving solution quality.

We tested the video pre-processing module on real-world videos. The videos were divided into four scenes based on traffic conditions (number of vehicles), and the performance of the module was evaluated by comparing with the module's output bounding boxes and speeds with annotations from humans according to the following metrics.
(i) \textbf{Average IoU}, defined as the mean intersection‑over‑union between every
manual bounding box and its best‑matching detection.  
(ii) \textbf{Precision}, measured by what fraction detected areas are real vehicles. (iii) \textbf{Recall},
calculated as the ratio between the number of detected vehicles and the total number of vehicles on the scene.
(iv) \textbf{F1 Score}, which is the precision and recall trade-off, and (v) \textbf{Speed Accuracy}, is the fraction of vehicles whose
per‑scene speed estimate deviates by less than 10\% from the
manually derived ground‑truth speed.
The results are summarized in Table \ref{tab:scene_metrics}.

\begin{table}[h]
    \centering
    \caption{Scene‑wise Detection and Speed‑Estimation Performance.}
    \label{tab:scene_metrics}
    \renewcommand{\arraystretch}{1.2}
    \begin{tabular}{c|ccccc}
        \hline
        \textbf{Scene} & \textbf{Avg IoU} & \textbf{Prec.} &
        \textbf{Recall} & \textbf{F1} & \textbf{Speed Acc.} \\
        \hline
        1 & 0.803 & 0.971 & 0.943 & 0.957 & 0.667 \\
        2 & 0.777 & 0.643 & 0.419 & 0.507 & 0.562 \\
        3 & 0.904 & 0.984 & 0.968 & 0.976 & 0.885 \\
        4 & 0.859 & 0.905 & 0.792 & 0.844 & 0.613 \\
        \hline
        \textbf{Average} & 0.836 & 0.876 & 0.781 & 0.821 & 0.682 \\
        \hline
    \end{tabular}
\end{table}

The module maintains a mean IoU of 0.836 and an F1‑score of 0.821,
while correctly estimating vehicle speeds within the 10\% error
tolerance in roughly 68.2\% of the cases, confirming robust performance
across diverse traffic scenes. While real-world video analysis validates our approach's feasibility, conducting real experiments in actual traffic environments presents significant challenges due to safety and cost constraints. To overcome these limitations, we conducted extensive experiments within the SUMO simulation environment for performance assessment across various traffic conditions.

\begin{figure}
    \centering
    \includegraphics[width=1\linewidth]{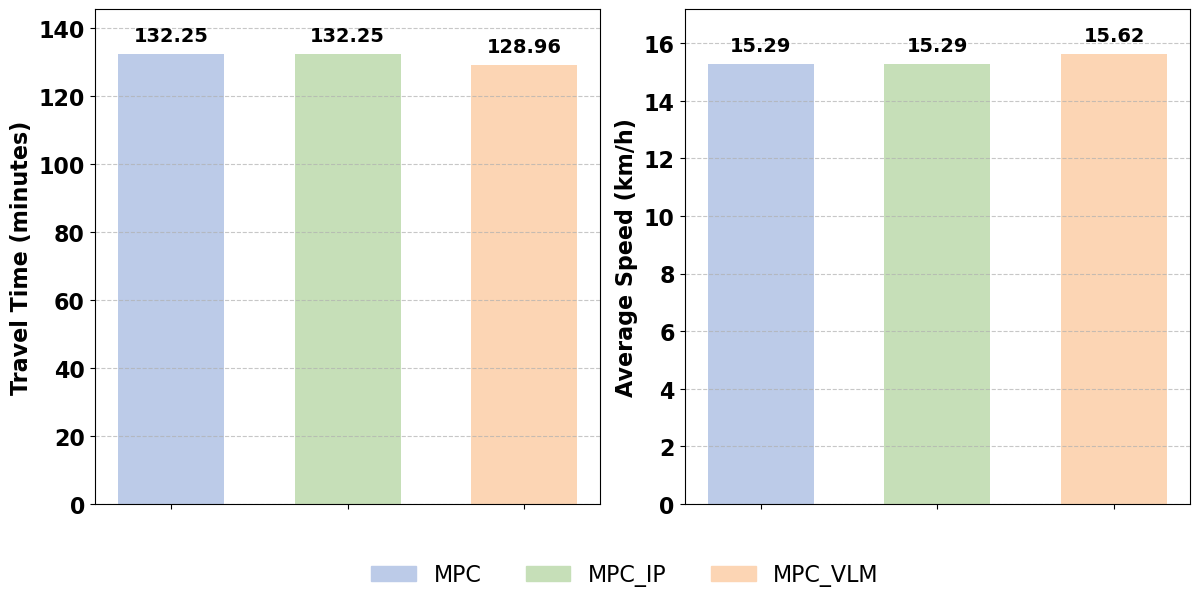}
    \caption{Travel efficiency comparison of different methods: MPC baseline, MPC\_IP, and MPC\_VLM}
    \label{fig:TRAVEL_EFF_medium}
\end{figure}

\subsection{Numerical Simulations}
We evaluated our proposed vision-language integrated trajectory planning framework against baseline approaches using several metrics: (i) \textbf{Travel Time/ Travel Speed}, measuring the average total duration required to complete a 2km journey, validated by corresponding average travel speed measurements; (ii) \textbf{Time Headway/ Distance Headway}, measuring the temporal and spatial gaps maintained between the ego vehicle and surrounding vehicles, with time headway representing the time needed to reach the position of leading vehicle and distance headway representing the physical distance, both critical for ensuring safe following behavior and collision avoidance; and (iii) \textbf{Collision Rate}, defined as the percentage of scenarios in which the output trajectories avoided collisions with obstacle vehicles. Collisions are identified when the safety verification layer detects bounding box intersections or insufficient safety margins as defined in Section IV.G.

We conducted comparative experiments evaluating three distinct approaches: the baseline method implementing only MPC with zero acceleration initialization; an enhanced version augmented with image processing capabilities (MPC\_IP) for obstacle detection; and our proposed VLM\_MPC framework.

As shown in Fig. \ref{fig:TRAVEL_EFF_medium}, which demonstrates the average travel times of all three approaches for the case of medium density, the MPC\_VLM approach achieved the shortest travel time of 128.96 seconds, outperforming the other two methods. Notably, MPC\_IP and the baseline MPC exhibit identical travel times and speeds. This equivalence occurs because our simulations were conducted in SUMO, which generates high-quality, noise-free bird's-eye view images with clear vehicle boundaries and optimal lighting conditions. Under these ideal visual conditions, the image processing pipeline achieves near-perfect object detection and tracking, extracting vehicle positions, speeds, and dimensions with accuracy equivalent to the ground-truth information directly available in the simulation environment. Meanwhile, SUMO already provides comprehensive vehicle information. Consequently, the image processing component in MPC\_IP does not contribute additional information beyond what is already available in the simulation environment, resulting in equivalent performance to the baseline MPC approach.

This result shows that the improved performance observed with MPC\_VLM do not come from superior object detection, but from the VLM's reasoning capabilities that enable enhanced trajectory initialization. 

We focus on medium traffic density scenarios for travel efficiency analysis as they provide reasonable conditions for evaluating trajectory planning performance. Low density conditions offer insufficient interaction complexity, while high density scenarios substantially limit the maneuvering options. High traffic density results are presented in the safety analysis to demonstrate robust collision avoidance performance under congested conditions.

\begin{table}[h]
    \centering
    \caption{Traffic Flow Comparison}
    \label{tab:traffic_flow}
    \renewcommand{\arraystretch}{1.2}
    \begin{tabular}{c|cc}
        \hline
        \textbf{Vehicle Type} & \textbf{Medium Density} & \textbf{High Density} \\
        & \textbf{(veh/hour)} & \textbf{(veh/hour)} \\
        \hline
        Medium Car & 2,400 & 2,800 \\
        Small Car & 600 & 800 \\
        Large Car & 400 & 600 \\
        Small Truck & 150 & 250 \\
        Large Truck & 50 & 80 \\
        \hline
        \textbf{Total} & \textbf{3,600} & \textbf{4,530} \\
        \hline
    \end{tabular}
\end{table}

\begin{figure}
    \centering
    \begin{subfigure}{\linewidth}
        \centering
        \includegraphics[width=\linewidth]{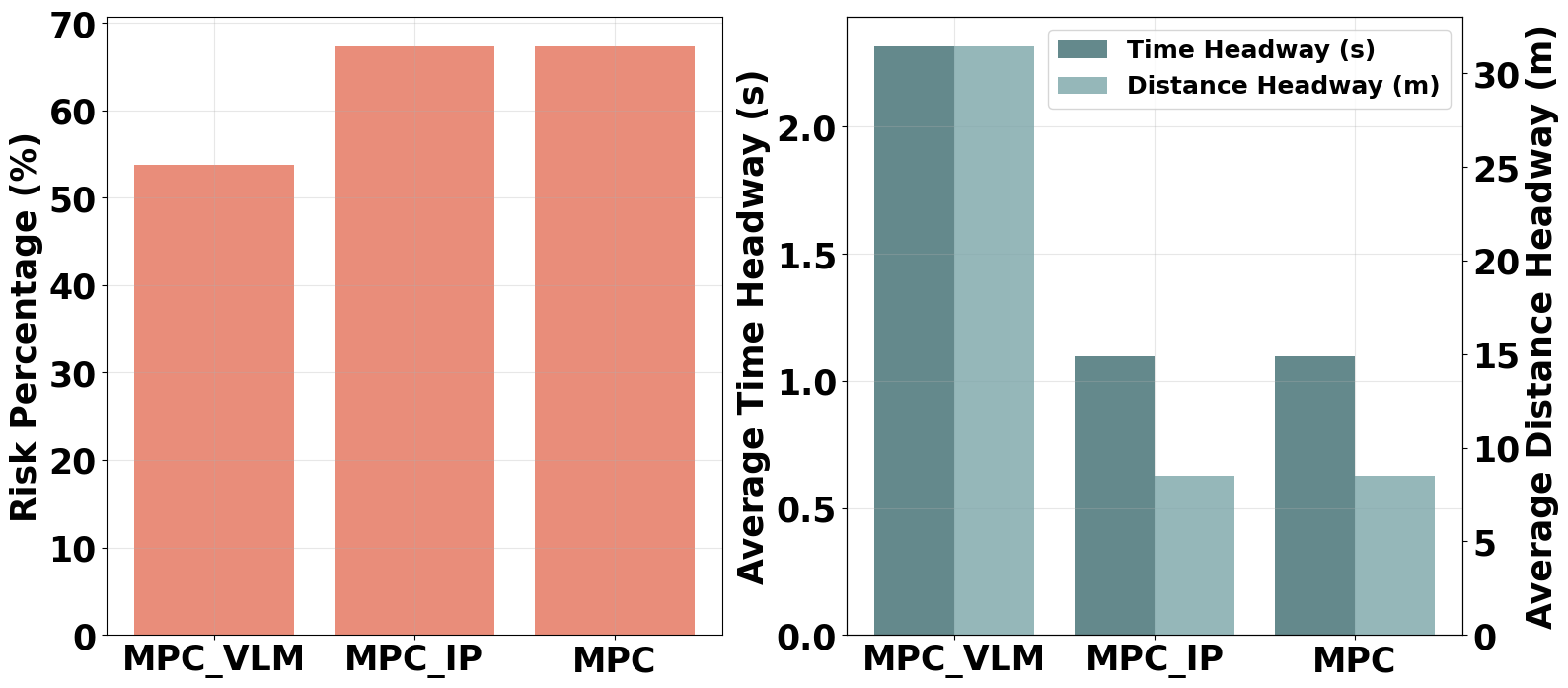}
        \caption{High traffic density}
        \label{fig:head_high}
    \end{subfigure}
    
    \vspace{0.5cm} 
    
    \begin{subfigure}{\linewidth}
        \centering
        \includegraphics[width=\linewidth]{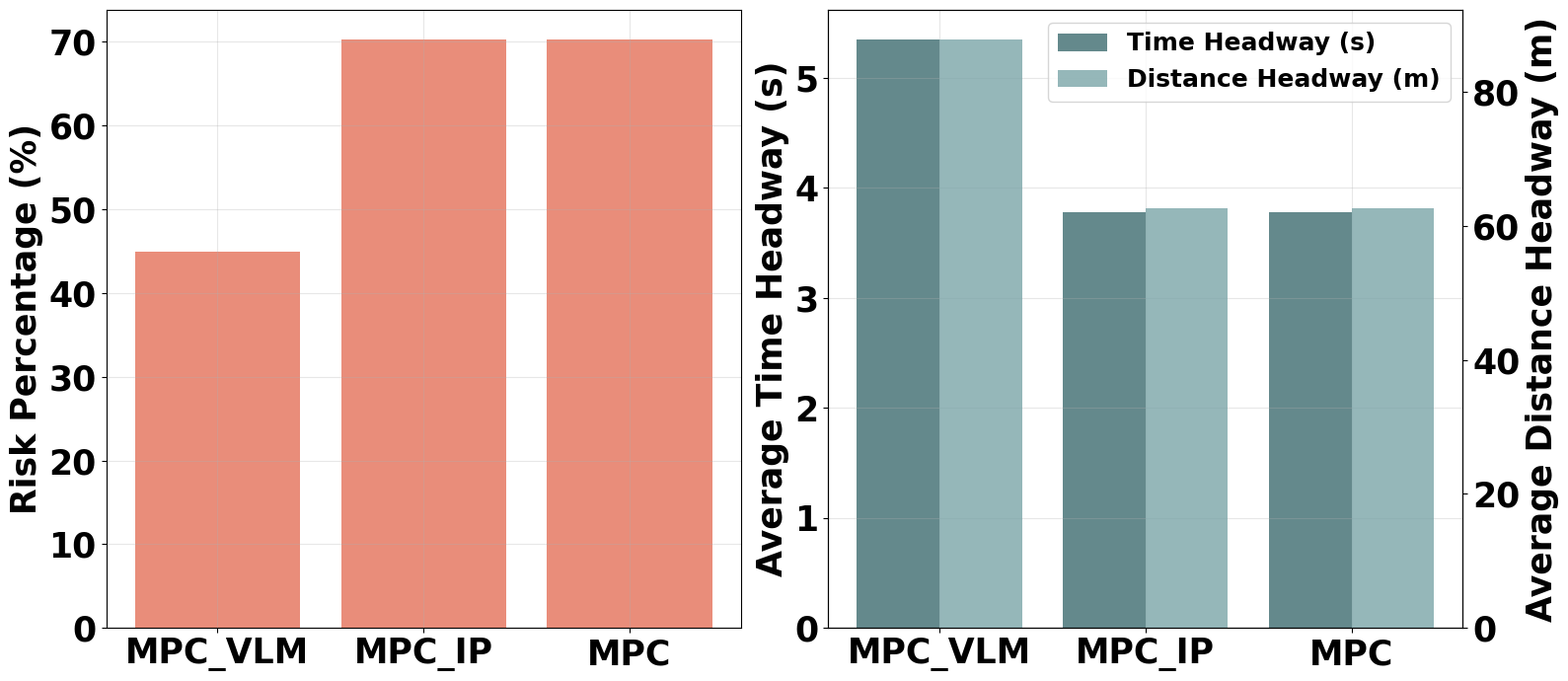}
        \caption{Medium traffic density}
        \label{fig:head_medium}
    \end{subfigure}
    
    \caption{Headway comparison among three methods under different traffic densities}
    \label{fig:headway_comparison}
\end{figure}

\begin{table*}[h]
\centering
\caption{Collision Rate Comparison}
\label{tab:coll_rate}
\begin{tabular*}{\textwidth}{@{\extracolsep{\fill}}c|c|c|c|c|c|c@{}}
\hline
{\multirow{3}{*}{Traffic Density}} & 
\multicolumn{3}{c|}{\textbf{Without Safety Layer}} & 
\multicolumn{3}{c}{\textbf{With Safety Layer}} \\
\cline{2-7}
& \shortstack{\textbf{Method}\\\textbf{name}} & \shortstack{\textbf{Collision}\\\textbf{Rate}} & \shortstack{\textbf{Average}\\\textbf{Dangerous Incidents}} & \shortstack{\textbf{Method}\\\textbf{name}} & \shortstack{\textbf{Collision}\\\textbf{Rate}} & \shortstack{\textbf{Average}\\\textbf{Dangerous Incidents}} \\
\hline
\multirow{3}{*}{Medium} & MPC Only & 0.286 & 173 & MPC Only & 0.000 & 73 \\
& MPC\_IP & 0.286 & 173 & MPC\_IP & 0.000 & 73 \\
& MPC\_VLM & 0.286 & 119 & MPC\_VLM & 0.000 & 22 \\
\hline
\multirow{3}{*}{High} & MPC Only & 0.286 & 233 & MPC Only & 0.000 & 68 \\
& MPC\_IP & 0.286 & 233 & MPC\_IP & 0.000 & 68 \\
& MPC\_VLM & 0.286 & 114 & MPC\_VLM & 0.000 & 27 \\
\hline
\end{tabular*}
\end{table*}

\subsection{Experimental Evaluations}
Fig. \ref{fig:headway_comparison} demonstrates the safety improvements achieved through VLM-enhanced trajectory planning by analyzing headway distances under different traffic density conditions. The results show that the proposed \textit{VisioPath} framework (MPC\_VLM) consistently maintains larger headway distances compared to baseline MPC and image-processing augmented MPC (MPC\_IP) approaches across both medium and high traffic densities. In medium density traffic, MPC\_VLM achieves an average headway of approximately 2.0 seconds compared to 1.5 seconds for baseline MPC, representing a 33\% improvement in safety margins. This enhancement is attributed to the VLM's ability to provide superior initial trajectory guidance that helps the DDP optimization converge to better local minima, enabling the ego vehicle to exploit available space in its surrounding environment more effectively. The improved awareness enables the autonomous vehicle to execute more strategic lane-changing maneuvers when safe opportunities arise, rather than remaining stuck behind slower vehicles in congested lanes. Under high traffic density conditions, where space is more constrained, the VLM-guided approach still maintains measurably larger headways, demonstrating its robustness in challenging scenarios. The improvements in headway across all different traffic conditions indicate that the VLM enhanced framework improves both efficiency and safety by maintaining appropriate distances from the surrounding vehicles and enabling more intelligent navigation decisions in mixed-traffic environments.

Table \ref{tab:coll_rate} presents an analysis of collision safety performance across different traffic densities, highlighting the effectiveness of our proposed safety verification layer. Without the safety verification layer, all three methods exhibit identical collision rates under both medium and high traffic densities, indicating that trajectory optimization alone is insufficient to guarantee collision-free operation in dense traffic scenarios. However, the introduction of the safety verification layer significantly improves the safety performance, and reduces collision rates substantially across all methods and traffic conditions.

Specifically, the results reveal that MPC\_VLM demonstrates superior safety capabilities even before safety verification is applied. Under medium traffic density, MPC\_VLM generates trajectories with 119 dangerous incidents compared to 173 for both baseline approaches, leading to a 31\% reduction in risky instances. This improvement is more profound under high traffic density, where MPC\_VLM results in only 114 dangerous incidents compared to 233 for the baseline methods, representing a 51\% reduction. After applying the safety verification layer, MPC\_VLM maintains its advantage with only 22 dangerous incidents in medium density (compared to 73 for other methods) and 27 incidents in high density (compared to 68 for other methods). These results indicate that the VLM-enhanced trajectory initialization leads to both improved travel efficiency and safe trajectory generation that require fewer safety interventions.

\subsection{Conclusion and Future Work}

In this work, we introduced \textit{VisioPath}, a novel framework that successfully integrates VLMs with MPC to enable safe autonomous navigation in mixed-traffic environments. Our approach demonstrates that combining the reasoning capabilities of modern VLMs with the mathematical rigor of DDP can achieve significant improvements over conventional MPC baselines. The framework leverages VLM-derived vehicle information to construct adaptive, elliptical-shaped potential field functions for collision avoidance that are seamlessly integrated into the optimization problem, while an event-triggered replanning strategy and real-time bird's-eye view preprocessing ensure computational efficiency. Additionally, VLM provides the initial trajectory guess for the DDP based on the surrounding traffic, which improves the convergence and solution quality of the algorithm. Through extensive simulations in SUMO, we demonstrated that \textit{VisioPath} outperforms baseline methods across multiple metrics, including travel efficiency, safety margins, and collision avoidance.

Future research directions include (1) real-world validation and deployment on physical autonomous vehicles; (2) enhanced VLM integration with multi-modal fusion techniques; (3) formal safety guarantee development through reachability analysis; (4) multi-agent coordination for improved traffic flow efficiency; and (5) adaptive learning capabilities for personalized driving behavior.

\bibliographystyle{IEEEtranS} 
\bibliography{myIEEEbib, IDS_Publications}

\begin{thebibliography}{10}
\providecommand{\url}[1]{#1}
\csname url@samestyle\endcsname
\providecommand{\newblock}{\relax}
\providecommand{\bibinfo}[2]{#2}
\providecommand{\BIBentrySTDinterwordspacing}{\spaceskip=0pt\relax}
\providecommand{\BIBentryALTinterwordstretchfactor}{4}
\providecommand{\BIBentryALTinterwordspacing}{\spaceskip=\fontdimen2\font plus
\BIBentryALTinterwordstretchfactor\fontdimen3\font minus \fontdimen4\font\relax}
\providecommand{\BIBforeignlanguage}[2]{{%
\expandafter\ifx\csname l@#1\endcsname\relax
\typeout{** WARNING: IEEEtranS.bst: No hyphenation pattern has been}%
\typeout{** loaded for the language `#1'. Using the pattern for}%
\typeout{** the default language instead.}%
\else
\language=\csname l@#1\endcsname
\fi
#2}}
\providecommand{\BIBdecl}{\relax}
\BIBdecl

\bibitem{alcan2022differential}
G.~Alcan and V.~Kyrki, ``Differential dynamic programming with nonlinear safety constraints under system uncertainties,'' \emph{IEEE Robotics and Automation Letters}, vol.~7, no.~2, pp. 1760--1767, 2022.

\bibitem{aradi2020survey}
S.~Aradi, ``Survey of deep reinforcement learning for motion planning of autonomous vehicles,'' \emph{IEEE Transactions on Intelligent Transportation Systems}, vol.~23, no.~2, pp. 740--759, 2020.

\bibitem{bochkovskiy2020yolov4}
A.~Bochkovskiy, C.-Y. Wang, and H.-Y.~M. Liao, ``Yolov4: Optimal speed and accuracy of object detection,'' \emph{arXiv preprint arXiv:2004.10934}, 2020.

\bibitem{choudhary2024talk2bev}
T.~Choudhary, V.~Dewangan, S.~Chandhok, S.~Priyadarshan, A.~Jain, A.~K. Singh, S.~Srivastava, K.~M. Jatavallabhula, and K.~M. Krishna, ``Talk2bev: Language-enhanced bird’s-eye view maps for autonomous driving,'' in \emph{2024 IEEE International Conference on Robotics and Automation (ICRA)}.\hskip 1em plus 0.5em minus 0.4em\relax IEEE, 2024, pp. 16\,345--16\,352.

\bibitem{fourati2024xlm}
S.~Fourati \emph{et~al.}, ``Xlm for autonomous driving systems: A comprehensive review,'' \emph{arXiv preprint arXiv:2409.10484}, 2024.

\bibitem{guanetti2018control}
J.~Guanetti, Y.~Kim, and F.~Borrelli, ``Control of connected and automated vehicles: State of the art and future challenges,'' \emph{Annual reviews in control}, vol.~45, pp. 18--40, 2018.

\bibitem{guo2024co}
Z.~Guo \emph{et~al.}, ``Co-driver: Vlm-based autonomous driving assistant with human-like behavior and understanding for complex road scenes,'' \emph{arXiv preprint arXiv:2405.05885}, 2024.

\bibitem{howell2022trajectory}
T.~A. Howell, S.~Le~Cleac’h, S.~Singh, P.~Florence, Z.~Manchester, and V.~Sindhwani, ``Trajectory optimization with optimization-based dynamics,'' \emph{IEEE Robotics and Automation Letters}, vol.~7, no.~3, pp. 6750--6757, 2022.

\bibitem{mayne1970}
D.~H. Jacobson and D.~Q. Mayne, \emph{Differential dynamic programming}, ser. Modern analytic and computational methods in science and mathematics.\hskip 1em plus 0.5em minus 0.4em\relax American Elsevier Pub. Co., 1970.

\bibitem{jahne2005digital}
B.~J{\"a}hne, \emph{Digital image processing}.\hskip 1em plus 0.5em minus 0.4em\relax Springer Science \& Business Media, 2005.

\bibitem{jallet2022constrained}
W.~Jallet, A.~Bambade, N.~Mansard, and J.~Carpentier, ``Constrained differential dynamic programming: A primal-dual augmented lagrangian approach,'' in \emph{2022 IEEE/RSJ International Conference on Intelligent Robots and Systems (IROS)}.\hskip 1em plus 0.5em minus 0.4em\relax IEEE, 2022, pp. 13\,371--13\,378.

\bibitem{kiran2021deep}
B.~R. Kiran, I.~Sobh, V.~Talpaert, P.~Mannion, A.~A. Al~Sallab, S.~Yogamani, and P.~P{\'e}rez, ``Deep reinforcement learning for autonomous driving: A survey,'' \emph{IEEE Transactions on Intelligent Transportation Systems}, vol.~23, no.~6, pp. 4909--4926, 2021.

\bibitem{Le2023ACC}
V.-A. Le and A.~A. Malikopoulos, ``Optimal weight adaptation of model predictive control for connected and automated vehicles in mixed traffic with bayesian optimization,'' in \emph{2023 American Control Conference (ACC)}.\hskip 1em plus 0.5em minus 0.4em\relax IEEE, 2023, pp. 1183--1188.

\bibitem{le2024controller}
------, ``{Controller Adaptation via Learning Solutions of Contextual Bayesian Optimization},'' \emph{arXiv preprint arXiv:2403.04881}, 2024.

\bibitem{li2025fine}
C.~Li, L.~Lux, A.~H. Berger, M.~J. Menten, M.~R. Sabuncu, and J.~C. Paetzold, ``Fine-tuning vision language models with graph-based knowledge for explainable medical image analysis,'' \emph{arXiv preprint arXiv:2503.09808}, 2025.

\bibitem{liu2021retinex}
S.~Liu, W.~Long, L.~He, Y.~Li, and W.~Ding, ``Retinex-based fast algorithm for low-light image enhancement,'' \emph{Entropy}, vol.~23, no.~6, p. 746, 2021.

\bibitem{mahbub2020sae-1}
A.~M.~I. Mahbub, V.~Karri, D.~Parikh, S.~Jade, and A.~A. Malikopoulos, ``A decentralized time- and energy-optimal control framework for connected automated vehicles: From simulation to field test,'' in \emph{SAE Technical Paper 2020-01-0579}.\hskip 1em plus 0.5em minus 0.4em\relax SAE International, 2020.

\bibitem{malikopoulos2021optimal}
A.~A. Malikopoulos, L.~Beaver, and I.~V. Chremos, ``Optimal time trajectory and coordination for connected and automated vehicles,'' \emph{Automatica}, vol. 125, p. 109469, 2021.

\bibitem{malikopoulos2018decentralized}
A.~A. Malikopoulos, C.~G. Cassandras, and Y.~J. Zhang, ``A decentralized energy-optimal control framework for connected automated vehicles at signal-free intersections,'' \emph{Automatica}, vol.~93, pp. 244--256, 2018.

\bibitem{murray1979constrained}
D.~M. Murray and S.~J. Yakowitz, ``Constrained differential dynamic programming and its application to multireservoir control,'' \emph{Water Resources Research}, vol.~15, no.~5, pp. 1017--1027, 1979.

\bibitem{murray1984differential}
------, ``Differential dynamic programming and {N}ewton's method for discrete optimal control problems,'' \emph{Journal of Optimization Theory and Applications}, vol.~43, no.~3, pp. 395--414, 1984.

\bibitem{nakka2022multi}
S.~K.~S. Nakka, B.~Chalaki, and A.~A. Malikopoulos, ``A multi-agent deep reinforcement learning coordination framework for connected and automated vehicles at merging roadways,'' in \emph{2022 American Control Conference (ACC)}.\hskip 1em plus 0.5em minus 0.4em\relax IEEE, 2022, pp. 3297--3302.

\bibitem{openai_o4}
{OpenAI}, ``{OpenAI o4-mini-high},'' \url{https://openai.com/index/introducing-o3-and-o4-mini/}, 2025, accessed: 2025-04-27.

\bibitem{philion2020lift}
J.~Philion and S.~Fidler, ``Lift, splat, shoot: Encoding images from arbitrary camera rigs by implicitly unprojecting to 3d,'' in \emph{Computer Vision--ECCV 2020: 16th European Conference, Glasgow, UK, August 23--28, 2020, Proceedings, Part XIV 16}.\hskip 1em plus 0.5em minus 0.4em\relax Springer, 2020, pp. 194--210.

\bibitem{rajamani2011vehicle}
R.~Rajamani, \emph{Vehicle dynamics and control}.\hskip 1em plus 0.5em minus 0.4em\relax Springer Science \& Business Media, 2011.

\bibitem{ren2015faster}
S.~Ren, K.~He, R.~Girshick, and J.~Sun, ``Faster r-cnn: Towards real-time object detection with region proposal networks,'' \emph{Advances in Neural Information Processing Systems}, vol.~28, 2015.

\bibitem{saravanos2023distributed}
A.~D. Saravanos, Y.~Aoyama, H.~Zhu, and E.~A. Theodorou, ``Distributed differential dynamic programming architectures for large-scale multiagent control,'' \emph{IEEE Transactions on Robotics}, 2023.

\bibitem{tassa2014control}
Y.~Tassa, N.~Mansard, and E.~Todorov, ``Control-limited differential dynamic programming,'' in \emph{2014 IEEE International Conference on Robotics and Automation (ICRA)}.\hskip 1em plus 0.5em minus 0.4em\relax IEEE, 2014, pp. 1168--1175.

\bibitem{tian2024drivevlm}
X.~Tian \emph{et~al.}, ``Drivevlm: The convergence of autonomous driving and large vision-language models,'' \emph{arXiv preprint arXiv:2402.12289}, 2024.

\bibitem{tomasi1998bilateral}
C.~Tomasi and R.~Manduchi, ``Bilateral filtering for gray and color images,'' in \emph{Sixth international conference on computer vision (IEEE Cat. No. 98CH36271)}.\hskip 1em plus 0.5em minus 0.4em\relax IEEE, 1998, pp. 839--846.

\bibitem{typaldos2023modified}
P.~Typaldos and M.~Papageorgiou, ``Modified dynamic programming algorithms for {GLOSA} systems with stochastic signal switching times,'' \emph{Transportation Research Part C: Emerging Technologies}, vol. 157, p. 104364, 2023.

\bibitem{typaldos2022optimization}
P.~Typaldos, M.~Papageorgiou, and I.~Papamichail, ``Optimization-based path-planning for connected and non-connected automated vehicles,'' \emph{Transportation Research Part C: Emerging Technologies}, vol. 134, p. 103487, 2022.

\bibitem{typaldos2020minimization}
P.~Typaldos, I.~Papamichail, and M.~Papageorgiou, ``Minimization of fuel consumption for vehicle trajectories,'' \emph{IEEE Transactions on Intelligent Transportation Systems}, vol.~21, no.~4, pp. 1716--1727, 2020.

\bibitem{wang2025corra}
S.~Wang, P.~Typaldos, and A.~A. Malikopoulos, ``Corra: Leveraging large language models for dynamic obstacle avoidance of autonomous vehicles,'' \emph{arXiv preprint arXiv:2503.02076}, 2025.

\bibitem{wei2022chain}
J.~Wei, X.~Wang, D.~Schuurmans, M.~Bosma, F.~Xia, E.~Chi, Q.~V. Le, D.~Zhou \emph{et~al.}, ``Chain-of-thought prompting elicits reasoning in large language models,'' \emph{Advances in Neural Information Processing Systems}, vol.~35, pp. 24\,824--24\,837, 2022.

\bibitem{xie2017differential}
Z.~Xie, C.~K. Liu, and K.~Hauser, ``Differential dynamic programming with nonlinear constraints,'' in \emph{2017 IEEE International Conference on Robotics and Automation (ICRA)}.\hskip 1em plus 0.5em minus 0.4em\relax IEEE, 2017, pp. 695--702.

\bibitem{xing2024autotrust}
S.~Xing \emph{et~al.}, ``Autotrust: Benchmarking trustworthiness in large vision language models for autonomous driving,'' \emph{arXiv preprint arXiv:2412.15206}, 2024.

\bibitem{yakowitz1986stagewise}
S.~Yakowitz, ``The stagewise {K}uhn-{T}ucker condition and differential dynamic programming,'' \emph{IEEE Transactions on Automatic Control}, vol.~31, no.~1, pp. 25--30, 1986.

\bibitem{yang2023llm4drive}
Z.~Yang, X.~Jia, H.~Li, and J.~Yan, ``Llm4drive: A survey of large language models for autonomous driving,'' \emph{arXiv preprint arXiv:2311.01043}, 2023.

\bibitem{zhou2024vision}
X.~Zhou, M.~Liu, E.~Yurtsever, B.~L. Zagar, W.~Zimmer, H.~Cao, and A.~C. Knoll, ``Vision language models in autonomous driving: A survey and outlook,'' \emph{IEEE Transactions on Intelligent Vehicles}, 2024.

\end{thebibliography}

\begin{IEEEbiography}[{\includegraphics[width=1in,height=1.25in,clip,keepaspectratio]{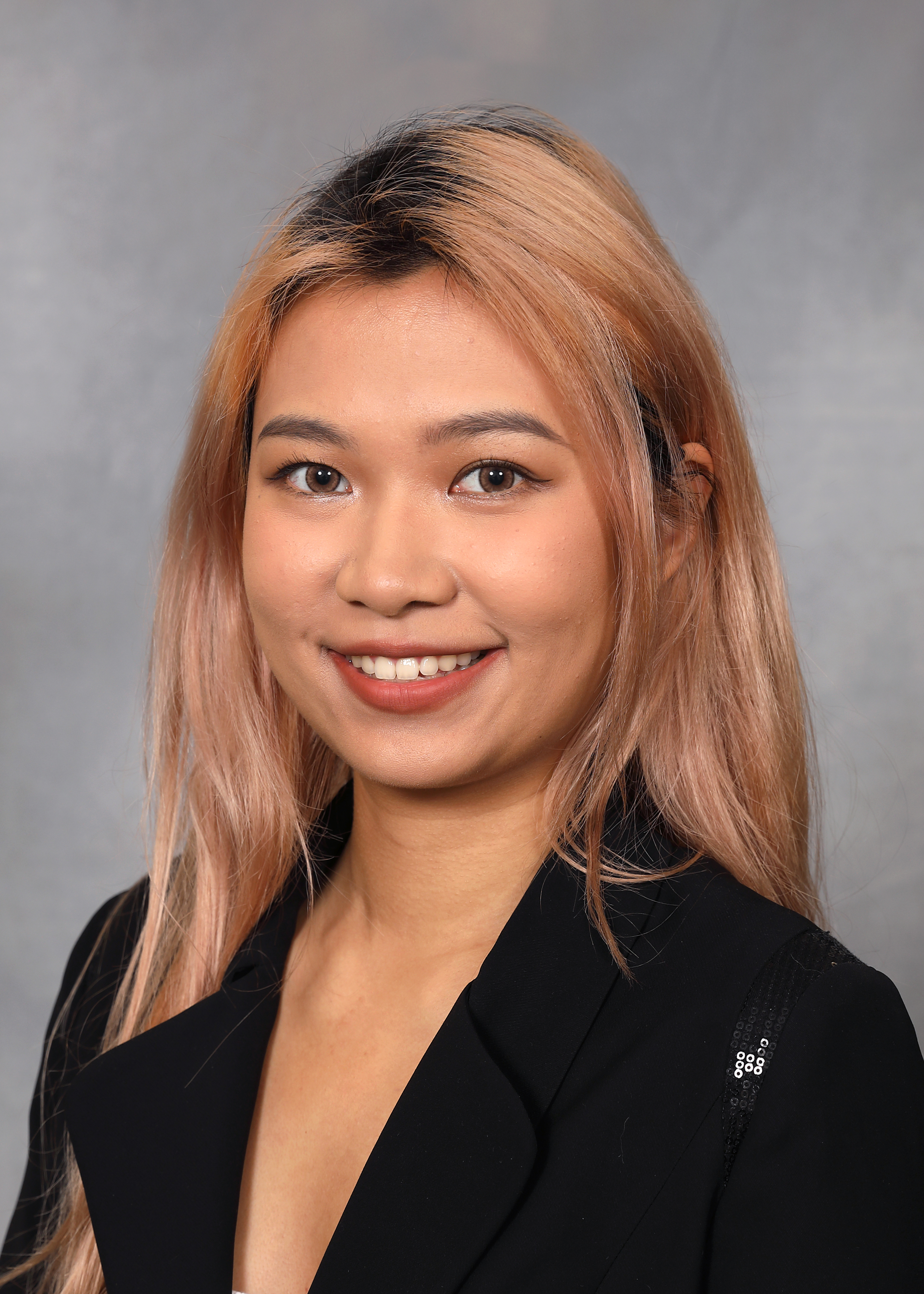}}]{Shanting Wang} {\space} received B.S. in Civil Engineering from University of Waterloo, ON, Canada and M.S. in System Engineering from Cornell University, Ithaca, NY, USA, in 2022 and 2024, respectively. She is currently continuing as a Ph.D student with System Engineering in Cornell University. Her research interests are on the integration of learning and decision-making in complex systems.
\end{IEEEbiography}

\begin{IEEEbiography}[{\includegraphics[width=1in,height=1.25in,clip,keepaspectratio]{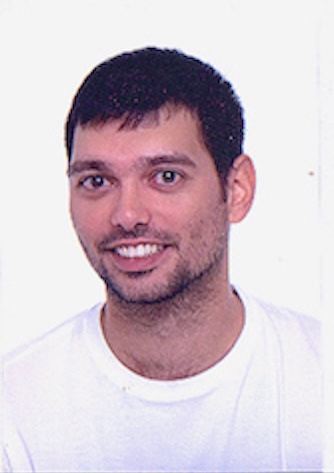}}]{Panagiotis Typaldos} {\space} received the B.S. degree in Applied Mathematics from the University of Crete, Heraklion, Greece in 2014 and the M.Sc. degree in Operational Research from the School of Production Engineering and Management, Technical University of Crete (TUC), Chania, Greece, in 2017. In 2022, Dr. Typaldos received his Ph.D. degree at TUC working on the effect of automated vehicles on highway traffic flow and signalized junctions. From May 2024, he is a postdoctoral researcher at the Information and Decision Science Laboratory (IDS) and a Visiting Instructor at Cornell University. His main research interests include optimal control and optimization theory and applications to intelligent transportation systems and trajectory planning for automated ground vehicles.
\end{IEEEbiography}

\begin{IEEEbiography}[{\includegraphics[width=1in,height=1.25in,
clip,keepaspectratio]{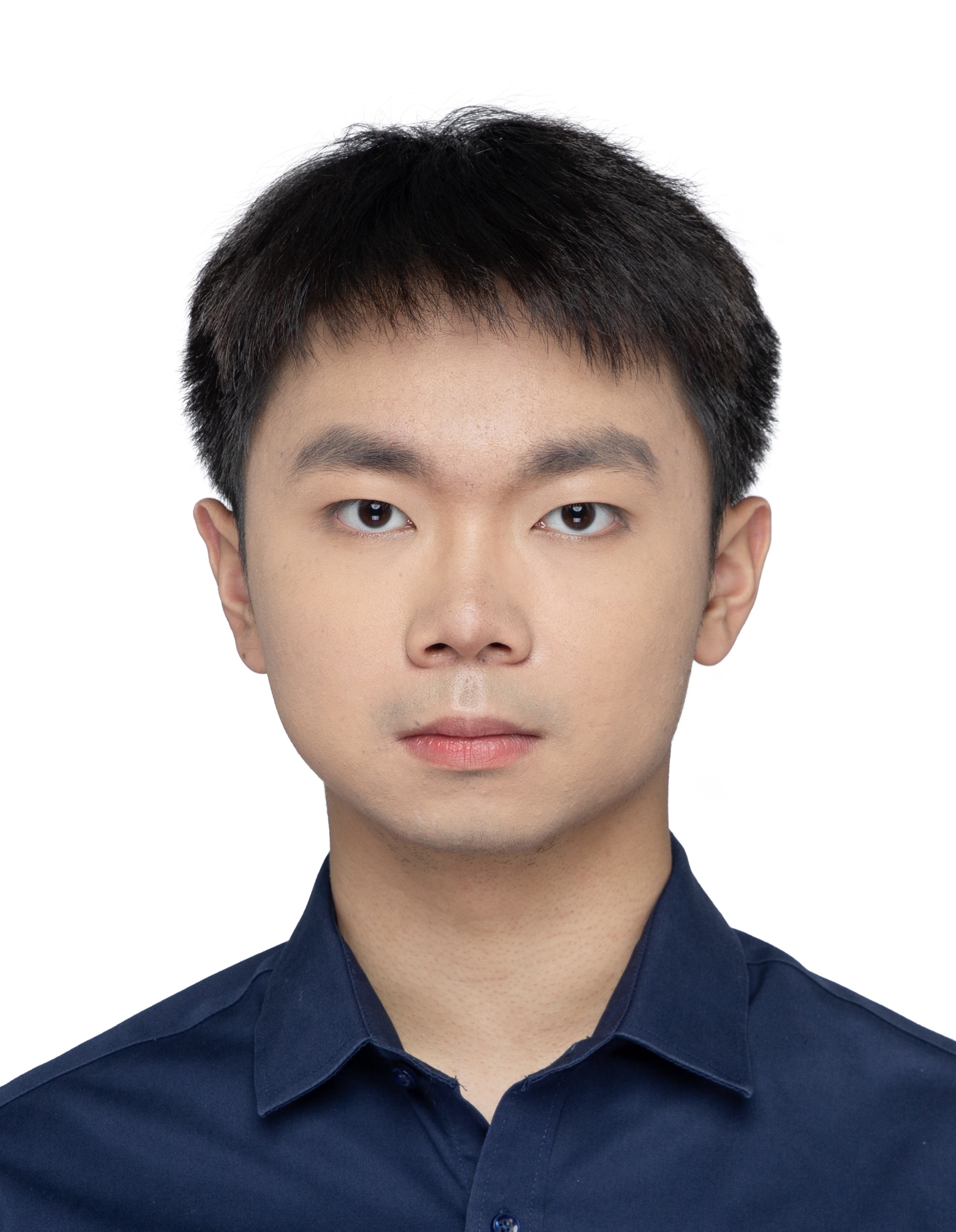}}]{Chenjun Li}{\space} received B.Eng. degrees in Electrical and Electronic Engineering from both University Glasgow, Glasgow, Scotland, and University of Electronic Science and Technology of China, Chengdu, China, in 2020. He is currently pursuing a Ph.D. degree in Electrical and Computer Engineering at Cornell University, Ithaca, NY, USA.
\end{IEEEbiography}

\begin{IEEEbiography}[{\includegraphics[width=1in,height=1.25in,clip,keepaspectratio]{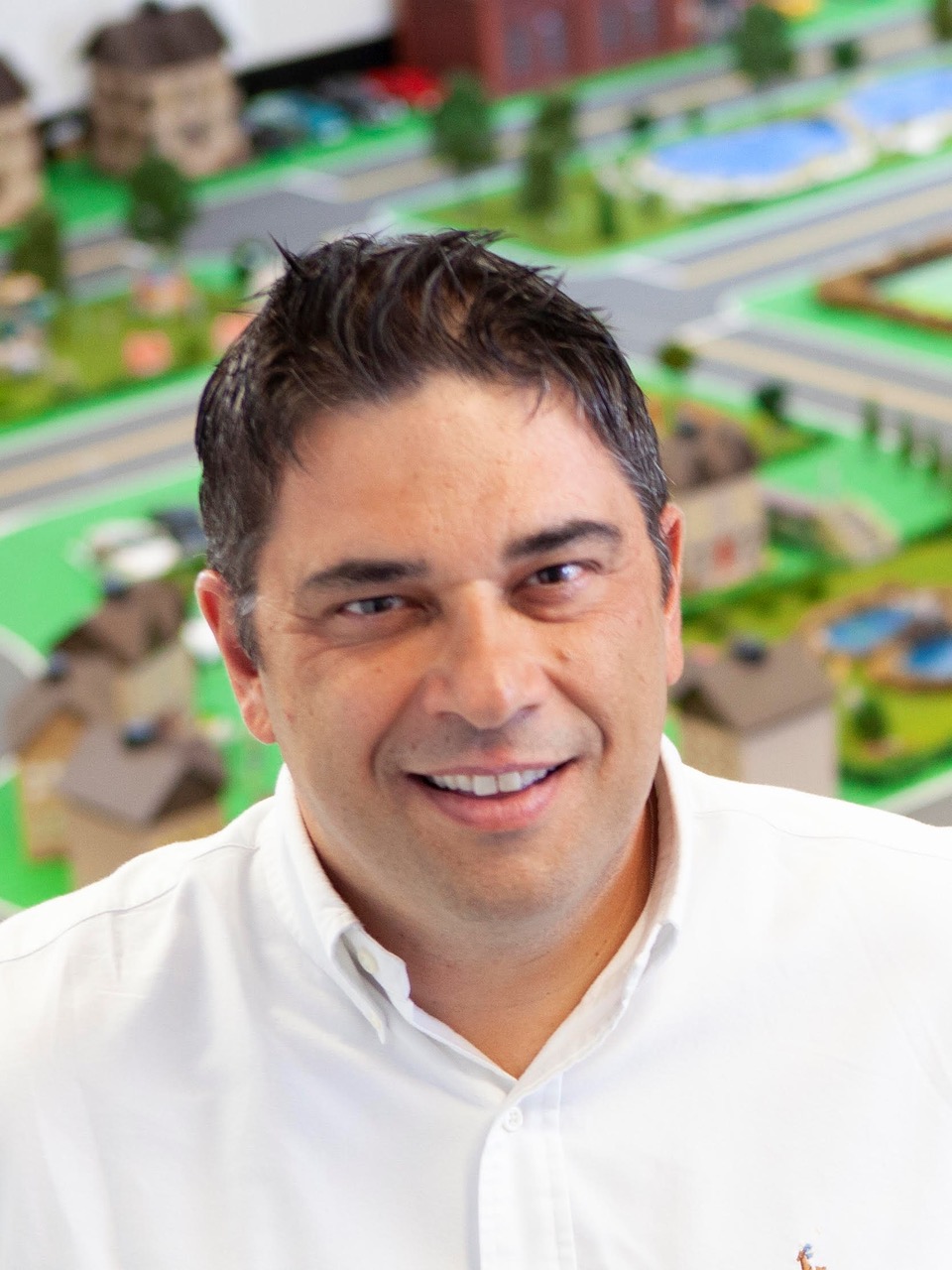}}]{Andreas A. Malikopoulos} (S’06–M’09–SM’17) {\space} received a Diploma in mechanical engineering from the National Technical University of Athens (NTUA), Greece, in 2000. He received M.S. and Ph.D. degrees in mechanical engineering at the University of Michigan, Ann Arbor, Michigan, USA, in 2004 and 2008, respectively. He is a professor at the School of Civil and Environmental Engineering at Cornell University and the director of the Information and Decision Science (IDS) Laboratory. Prior to these appointments, he was the Terri Connor Kelly and John Kelly Career Development Professor in the Department of Mechanical Engineering (2017-2023) and the founding Director of the Sociotechnical Systems Center (2019- 2023) at the University of Delaware (UD). Before he joined UD, he was the Alvin M. Weinberg Fellow (2010-2017) in the Energy \& Transportation Science Division at Oak Ridge National Laboratory (ORNL), the Deputy Director of the Urban Dynamics Institute (2014-2017) at ORNL, and a Senior Researcher in General Motors Global Research \& Development (2008-2010). His research spans several fields, including analysis, optimization, and control of cyber-physical systems (CPS); decentralized stochastic systems; stochastic scheduling and resource allocation; and learning in complex systems. His research aims to develop theories and data-driven system approaches at the intersection of learning and control for making CPS able to realize their optimal operation while interacting with their environment. He has been an Associate Editor of the IEEE Transactions on Intelligent Vehicles and IEEE Transactions on Intelligent Transportation Systems from 2017 through 2020. He is currently an Associate Editor of Automatica and IEEE Transactions on Automatic Control, and a Senior Editor of IEEE Transactions on Intelligent Transportation Systems. He is a member of SIAM, AAAS, and a Fellow of the ASME.
\end{IEEEbiography}

\end{document}